\documentclass{jfm}
\usepackage{graphicx}
\usepackage{subcaption}
\usepackage{amsmath}
\usepackage{epstopdf, epsfig}
\usepackage{color}
\usepackage{chngcntr}
\usepackage{physics}
\usepackage{mathtools}
 \usepackage{booktabs}

\usepackage{lineno}

\usepackage{url}
\usepackage[colorlinks=true,citecolor=blue,linkcolor=blue,urlcolor=blue]{hyperref}

\usepackage{xcolor}

\shorttitle{Nanoscale hydrodynamics of contact lines on high-friction surfaces}
\shortauthor{M. Pellegrino, P. K. Kannan et al.}

\title{Cahn-Hilliard Phase Field modelling captures nanoscale contact line dynamics on high-friction surfaces}

\author{Michele Pellegrino\aff{1,3}\corresp{\email{michele.pellegrino@polito.it}}\thanks{These authors contributed equally to this work},
    Parvathy K. Kannan\aff{2}\footnotemark[2],
    Gustav Amberg\aff{2},
    Shervin Bagheri\aff{2},
    Outi Tammisola\aff{2}
    \and Berk Hess\aff{1}
}

\affiliation{
    \aff{1}Swedish e-Science Research Centre, Science for Life Laboratory, Department of Applied Physics KTH, 10044 Stockholm, Sweden
    \aff{2}FLOW, Department of Engineering Mechanics, KTH Royal Institute of Technology, 11428 Stockholm, Sweden
    \aff{3}Department of Energy, Politecnico di Torino, 10129 Turin, Italy
}

\begin{document}

\maketitle

\begin{abstract}
Incorporating molecular-scale effects in the description of contact line motion is essential for accurately capturing all sources of energy dissipation in wetting dynamics. This holds particularly true in the cases where contact line friction dominates, and hydrodynamics models struggle to achieve regularisation due to the negligible Navier slip. We perform Molecular Dynamics simulations of water/hexane biphasic systems in a two-phase Couette flow configuration. Wetting occurs over a silica-like surface with controllable wettability. The simulation results are reproduced by a Phase Field model (Cahn-Hilliard Navier-Stokes equations), which includes localised contact line slip and contact angle dynamics. The continuous equations are directly parametrized from Molecular Dynamics simulation results, under the numerical sharp interface limit. We demonstrate that the Phase Field model can quantitatively reproduce Molecular Dynamics through a systematic calibration protocol. Critically, we show that contact line friction is the primary physical parameter requiring empirical calibration based on Molecular Dynamics data. Once extracted by matching contact angle dynamics, quantitative agreement across multiple observables is obtained, including interface curvature, steady contact line displacement, and the structure of streamlines. All other model parameters are determined \textit{a posteriori}, according to the calculation of independent observables and under numerical constraints. The results presented in this article indicate that Phase Field modelling can capture the net effect of molecular processes on the mobility of contact lines and that the careful calibration of contact line friction based on the reconstruction of contact angle dynamics and interface bending is key to fully reconcile continuous models with Molecular Dynamics.
\end{abstract}


\section{Introduction}
\label{sec:intro}

Accurate mesoscale models of wetting processes are of prime importance for understanding and controlling fluid dynamics from the millimetre down to the nanometre scale. Applications in this range of scales include additive manufacturing \citep{deruiter2018additive}, hydrogen storage \citep{vanrooijen2022hydrogen}, heat transfer \citep{bures2021boiling} and energy production \citep{liu2024nanosponge}. Wetting dynamics is determined by the motion of three-phase contact lines, i.e. the regions of space where an interface between fluids meets a solid surface. Despite moving contact lines being commonly observed, even in daily experience, their hydrodynamic description is not trivial. \cite{huh1970hydrodynamic} found that the solution of Stokes equations for the flow near a planar interface impinging on a flat surface presents a non-integrable stress singularity at the contact line, if no-slip boundary conditions are applied at the solid wall. Replacing the no-slip boundary conditions with Navier slip conditions relaxes the non-integrable stress singularity, as shown by \cite{dussan1976slip} and \cite{hocking1977slip}. The Navier slip model is however insufficient to completely regularize contact line dynamics, as it produces an integrable pressure divergence \citep{huh1977capillary}; this singularity can be avoided by allowing the contact angle to be a dynamic quantity \citep{fricke2019angle}, or by introducing a second-order slip condition \citep{kulkarni2023stream}.

Significant deviations from the no-slip boundary condition have been observed in Molecular Dynamics (MD) simulations, as already shown in the seminal works by \cite{koplik1988poiseuille}, \cite{koplik1989md}, and \cite{thompson1989slip}. \cite{ren2007mdslip} showed a substantial increase of the slip length in the contact line region for a nanoconfined two-phase Couette flow. Furthermore, experimental evidence \citep{deng2016experimental,ludwicki2022mobility} suggests that the contact angle between the liquid/liquid interface and the solid surface is a dynamic observable, whose deviation from equilibrium correlates with the velocity of the contact line. 

Liquid/solid slip, possibly localised in the contact line region and associated with contact angle dynamics, is therefore an essential ingredient required to model moving contact lines. Albeit micrometre-scale slip lengths have been measured on patterned or lubricated surfaces, the slip length on flat hydrophilic surfaces is usually on the order of nanometers, sometimes too small to even be measured reliably \citep{lauga2005microfluidics}. Very small slip lengths represent a challenge for the computational modelling of contact lines, since they entail unattainable mesh refinement in the contact line region \citep{afkhami2009vof}. Therefore, most common approaches in computational fluid dynamics adopt slip lengths that are unphysically large, simply to ensure numerical feasibility. While this scale-separation challenge persists, the flow and wetting community has reached no consensus regarding the physical validity of macroscale models \citep{shikhmurzaev2020test,bothe2020reflections,shikhmurzaev2020reflections2}. This point stresses the importance of accurate mesoscale models that can serve as an intermediary between the molecular and macroscopic descriptions of contact lines, as highlighted in the review by \cite{smith2018review}.  Among several continuous models, Phase Field techniques have proven remarkably effective in reproducing the motion of contact lines observed in experiments \citep{carlson2011dissipation} and molecular simulations \citep{qian2003gnbc,diewald2020gradient}. A popular `flavour' of Phase Field models are Cahn-Hilliard-Navier-Stokes equations (CHNS), conceived to reproduce the dynamics of interfaces between fluids in the absence of phase transitions \citep{jacqmin1999chns}. The success of CHNS equations originates from their capability to displace contact lines by virtue of a localised \textit{diffusive slip} owing to a finite interface thickness, which is able to regularize the stress singularity even in the absence of Navier slip. Furthermore, contact angle dynamics can be easily incorporated into the boundary conditions of CH equations, as shown by \cite{yue2011wallrelax}.

Several experimental techniques have been used to verify hydrodynamic models of contact lines. Traditional techniques such as particle image velocimetry or goniometry have greatly evolved and can now provide abundant information on the interface reconfiguration and the flow profile in wetting dynamics \citep{xia2020oscillatory,gupta2024experimental}. However, experimental approaches relying on optical microscopy cannot resolve the nanoscale structure of flow and interfaces close to contact lines, as it lies below the optical resolution limit. On the other hand, techniques capable of probing nanoscopic physics, such as atomic force microscopy \citep{deng2017afm}, cannot provide a detailed picture of the fluid dynamics occurring at those scales. Due to these experimental limitations, MD simulations have emerged as a complementary tool that can be used to reach the nanoscopic scale, while providing very detailed information on the fluid dynamics of contact lines \citep{fernandeztoledano2020md,giri2022microscopic,badr2024polymer}. At the same time, MD simulations allow to finely control both the wettability of the surface and the speed of the wetting process. Furthermore, one can obtain both large and small slip length, depending on the geometry and the chemical parametrization of the solid surface. In this work, we utilise a silica-like surface that yields \textit{virtually} zero slip with water: this choice allows us to test the capability of CHNS to introduce slip only locally to the contact line.

The viscosity of the wetting fluids and the friction between the fluids and the solid surface in the contact line region, contact line friction, constitute the two channels of energy dissipation in wetting dynamics on flat rigid surfaces. Viscosity and friction affect the curvature of the fluid-fluid interface, giving rise to a deviation from the equilibrium configurations determined by the combination of Young-Laplace and Young-Dupr\'{e} equations. In particular, viscous dissipation bends the interface proportionally to the flow velocity gradient, while contact line friction tilts the contact angle proportionally to the speed of the contact line. \cite{doquang2015taxonomy} suggested that the effect of contact line friction may be substantial if wetting occurs on a high-energy surface and the viscosity of the wetting liquid is sufficiently small. 

The most convenient system used in MD simulations to study wetting dynamics is a two-phase Couette flow: a liquid meniscus is confined between two solid walls, with vapour forming the other fluid phase. This setup is inconvenient when analysing the flow profile in the proximity of contact lines, since density and velocity fields cannot be sampled in the vapour phase for a nanoconfined system, where the Knudsen number is $O(1)$. We instead simulate a system composed by two immiscible liquids, water and hexane, which can be more readily compared to the continuous counterpart. It is known that streamlines resulting from CHNS simulations may cross the liquid/liquid interface close to the contact line, as a result of the local diffusive slip \citep{lacis2020johansson}. We therefore compare the flow profiles between MD and CHNS and study whether streamline crossing may be mitigated in CHNS simulations by choosing a \textit{mobility} parameter that is as tight as possible to the \textit{sharp interface} condition.

The main goal of this work is to reproduce interface curvature quantitatively and for a wide range of typical flow velocities, therefore testing if CHNS can accurately capture both sources of energy dissipation. This is achieved by establishing a systematic protocol requiring minimal parameter fitting. We demonstrate that contact line friction coefficients for advancing and receding contact lines are the only parameters required to be calibrated against MD. Once these are determined, Phase Field parameters (interface thickness and mobility) are set primarily by numerical considerations, namely the sharp interface condition. Shear viscosity coefficients are measured independently from MD, and a sub-nanometric Navier slip is introduced only to ensure numerical stability. This protocol yields quantitative reproduction of interface curvature for a wide range of contact line speeds, validating that contact line friction is the essential molecular input for accurate mesoscale wetting models. We further test this calibration by comparing steady contact line displacement and flow structure between MD and CHNS, and examine the limits of validity as velocity increases.

The paper is organised as follows. In Section~\ref{sec:chns} we introduce Cahn-Hilliard Navier-Stokes equations, with a focus on the wetting, boundary and sharp interface conditions. In Section~\ref{sec:md-simulations} we describe the system simulated using Molecular Dynamics, how fluid flow is driven and how interfaces and contact line friction are determined. In Section~\ref{sec:md-vs-pf} we compare the results of the two simulation techniques, and in particular interface curvature, steady displacement of contact lines and flow profiles. In Section~\ref{sec:discussion} we discuss the sensitivity of numerical and physical parameters and the attainment of the sharp interface limit. Finally, in Section~\ref{sec:conclusions} we draw the conclusions and highlight the advancement with respect to previous work.

\section{Cahn-Hilliard Navier-Stokes equations} 
\label{sec:chns}

\subsection{Equations and boundary conditions}  \label{sec:chns-formulation}

\begin{figure}
    \centering
    \includegraphics[width=0.8\linewidth,trim={0 2cm 0 2cm},clip]{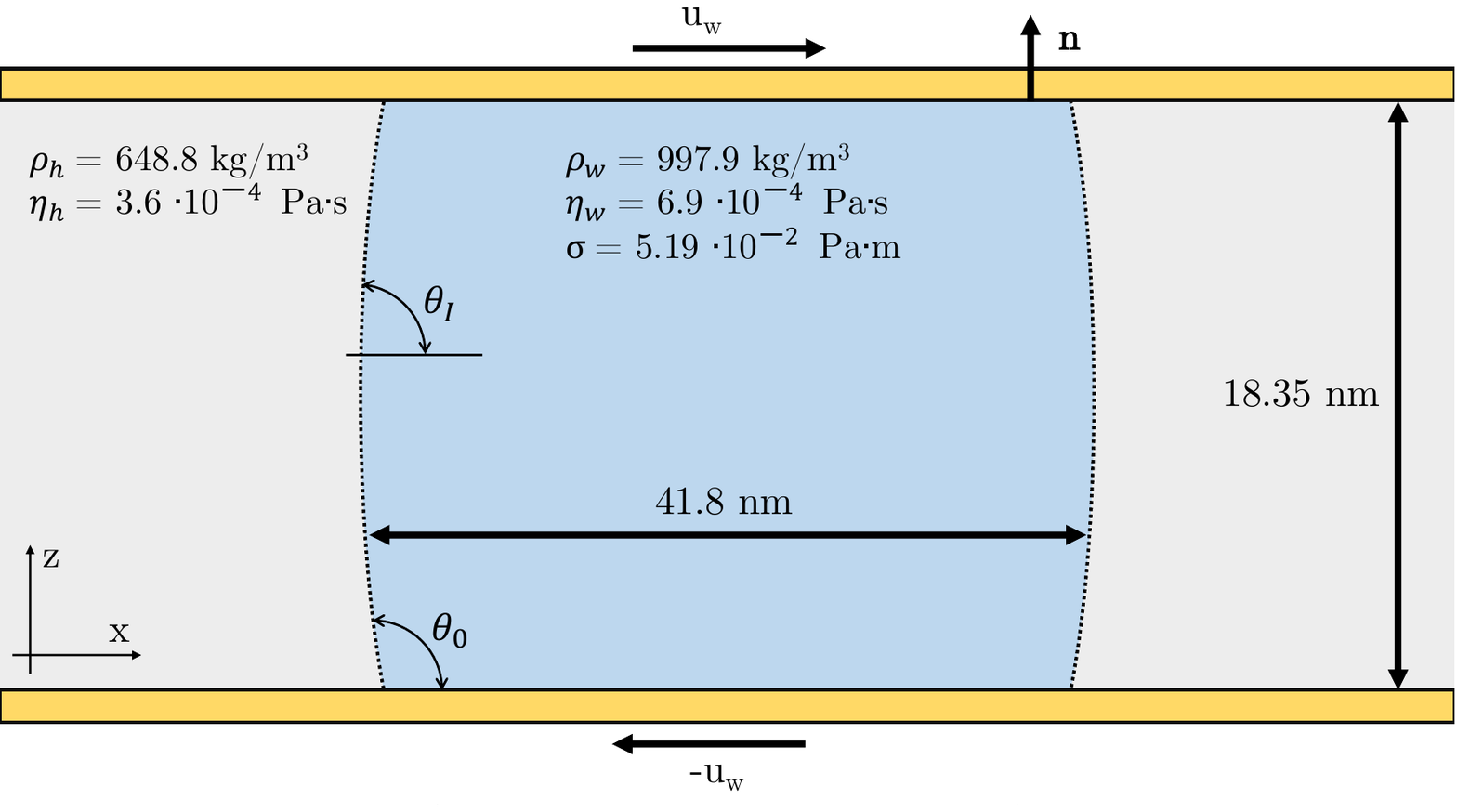}
    \caption{Geometry of the system simulated with CHNS equations. Material properties are obtained from MD simulations.}
    \label{fig:chns-geometry}
\end{figure}

The simulated biphasic system is sketched in Figure~\ref{fig:chns-geometry}. The Phase Field model defines the free energy per unit volume of the binary mixture \citep{cahn1958ch} as follows:
\begin{equation}    \label{eq:free-energy}
    \mathcal{F} = \frac{3}{4\sqrt{2}}\varepsilon  \sigma |\bnabla C|^2 + \frac{3}{2\sqrt{2}}\frac{\sigma}{\varepsilon} \psi (C) ,
\end{equation}
$\varepsilon$ being the interface thickness, $C$ the order parameter ranging from $-1$ in hexane to $+1$ in water and $\sigma$ the water/hexane surface tension, measured from MD to be $5.19\times10^{-2}$ Pa$\cdot$m (see appendix \ref{sec:app-md}). The double well potential function $\psi = (C^2 -1)^2/4$ has minima in the bulk phases. The first term denotes the interfacial energy making the components of the binary system want to mix, whereas the latter denotes the bulk energy that prefers total phase separation of the two phases into domains of pure components.

The motion of the diffuse interface in the form of the transport of the order parameter is governed by Cahn-Hilliard equation:
\begin{equation}    \label{eq:cahn-hilliard}
    \frac{\partial C}{\partial t} + \textbf{u}\cdot\bnabla C = \bnabla\cdot[M \bnabla \phi] \; ,
\end{equation}
with
\begin{equation}    \label{eq:chemical-potential}
    \phi = \frac{3}{2\sqrt{2}} \frac{\sigma}{\varepsilon} \psi ' (C) - \frac{3}{2\sqrt{2}} \sigma \varepsilon \bnabla ^2 C
\end{equation}
being the chemical potential derived by taking the variational derivative of the free energy equation. A change in chemical potential across the interface drives it forward. $M$ is the mobility parameter which we set by using the sharp interface condition found to be valid for no-slip boundary conditions \citep{yue2011diffcontlines}. This will be discussed in detail in Section~\ref{sec:chns-sil}. 

In order to couple the Cahn-Hilliard equations to the Navier-Stokes equations, a surface tension source term is added to the Navier-Stokes which results in the following CHNS equations:
\begin{equation}    \label{eq:ch-ns}
    \begin{dcases}
        \rho(C) \left(\frac{\partial \textbf{u}}{\partial t} + \textbf{u}\cdot\bnabla \textbf{u}\right) = -\bnabla p + \bnabla\left[\eta(C)\left(\bnabla\textbf{u} + \bnabla\textbf{u}^T\right)\right] + \phi\bnabla C \\
        \bnabla\cdot\textbf{u} = 0
    \end{dcases} \; ,
\end{equation}
with $\rho(C)$ and $\eta(C)$ being respectively the density and the viscosity of the two liquid phases. Note that there is no gravity contribution in Equation~\ref{eq:ch-ns}, owing to the size of the system being much smaller than the capillary length. These equations are solved using the following boundary conditions at the walls:
\begin{equation}    \label{eq:nav-slip}
    u_x = u_w - \ell_N\frac{\partial u_x}{\partial \boldsymbol{n}} \; , \quad u_z=0 \; ,
\end{equation}
\begin{equation}    \label{eq:diff-flux}
    \textbf{n}\cdot\bnabla \phi = 0 \; ,
\end{equation}
\begin{equation}    \label{eq:wet-bc}
    \frac{\partial C}{\partial t} + \textbf{u}\cdot\bnabla C = -\Gamma \left( \frac{3 }{2\sqrt{2}} \textbf{n}\cdot\bnabla C + \frac{3}{4} \sigma(1-C^2)\cos\theta_0 \right) \; ,
\end{equation}
where $\Gamma$ is a positive phenomenological parameter that determines how fast the contact line relaxes to the equilibrium angle $\theta_0$. $\ell_N$ is a sub-nanometric Navier slip length introduced as numerical regularization. 

We emphasise that Equation~\ref{eq:wet-bc} is not an \textit{ad hoc} closure but follows from the variational structure of the model. Supplementing the free energy~\ref{eq:free-energy} with a surface free-energy density that depends on the boundary value of the concentration, and taking its variational derivative with respect to $C$ at the wall, yields an equilibrium wetting condition in which $\theta_0$ enters through the solid--fluid interfacial energies \citep{jacqmin1999chns,yue2011wallrelax}. Equation~\ref{eq:wet-bc} relaxes this condition dynamically, with $\Gamma$ setting the rate at which the contact line approaches the wall's preferred energy state.

The combination of a thin diffuse interface and zero-slip boundary conditions creates an extremely stiff numerical problem requiring very small time steps and leading to convergence difficulties. Finite slip, kept physically negligible (Table \ref{tab:cl-steady-param-match}), ensures numerical stability. Equations~\ref{eq:nav-slip} and \ref{eq:diff-flux} are the Navier-slip, impermeability and the zero diffusive flux boundary condition, whereas Equation~\ref{eq:wet-bc} is the wall-energy relaxation boundary condition which governs the dynamics of the contact line according to the contact angle \citep{carlson2011dissipation}. A contact line friction coefficient can be defined by comparing results by \cite{yue2011wallrelax} on the wetting boundary condition for a steady slow flow with Molecular Kinetic Theory \citep{blake1969mkt}:
\begin{equation}    \label{eq:muf-def-pf}
    \mu_f = \frac{1}{\varepsilon\Gamma} \; .
\end{equation}
The relaxation parameter $\Gamma$ has dimensions of $(\text{Pa}\cdot\text{s}\cdot\text{m})^{-1}$. In the simulations we only calibrate $\mu_f$, holding the interface thickness $\varepsilon$ to a fixed value, bounded from below by the MD interface width and large enough that the interface can be resolved with a computationally feasible mesh (see Section~\ref{sec:chns-sil}). The relaxation parameter $\Gamma$ then follows from $\mu_f$ through Equation~\ref{eq:muf-def-pf}, rather than being tuned independently. For $\varepsilon = 0.3$~nm and $\mu_f$ of Table~\ref{tab:cl-steady-param-match}, $\Gamma$ lies in the range $\approx 2.4\times10^{11}$--$1.2\times10^{12}~(\text{Pa}\cdot\text{s}\cdot\text{m})^{-1}$.


The full system of fluid and Phase Field equations, together with the boundary conditions, is discretised and solved using open-source code FreeFEM \citep{hecht2012freefem}. Adaptive mesh is used near the two-phase interface to capture the variation of the function $C$. Additional details on the weak formulation of the CHNS equations and on the numerical methods are provided in the supplementary information.

\subsection{Interface sharpness}  \label{sec:chns-sil}

In order to select the appropriate range of free parameters, $\varepsilon$, $M$ and $\ell_N$, to match the MD simulations, we start by fixing a physically sensible interface thickness $\varepsilon$. Due to numerical limitations, the interface thickness of CHNS equations typically overestimates the intrinsic interface width of immiscible molecular systems. In our case, we use $\varepsilon=0.3$ nm. Next, the range of the mobility parameter $M$ is obtained by considering the sharp interface condition found by \cite{yue2010sil} as shown below:
\begin{equation}    \label{eq:sharp-interface-limit}
    \varepsilon_{md} \le \varepsilon \le 4\sqrt{M\eta^*} \; .
\end{equation}
being $\eta^*=\sqrt{\eta_w \eta_h}$ the effective viscosity at the interface, defined as the geometric average of the viscosity coefficients of the two liquids.

\section{Molecular Dynamics simulations}
\label{sec:md-simulations}

\subsection{Force fields and configuration}
\begin{figure}
    \centering
    \includegraphics[width=0.95\linewidth,trim={0 2cm 0 2cm},clip]{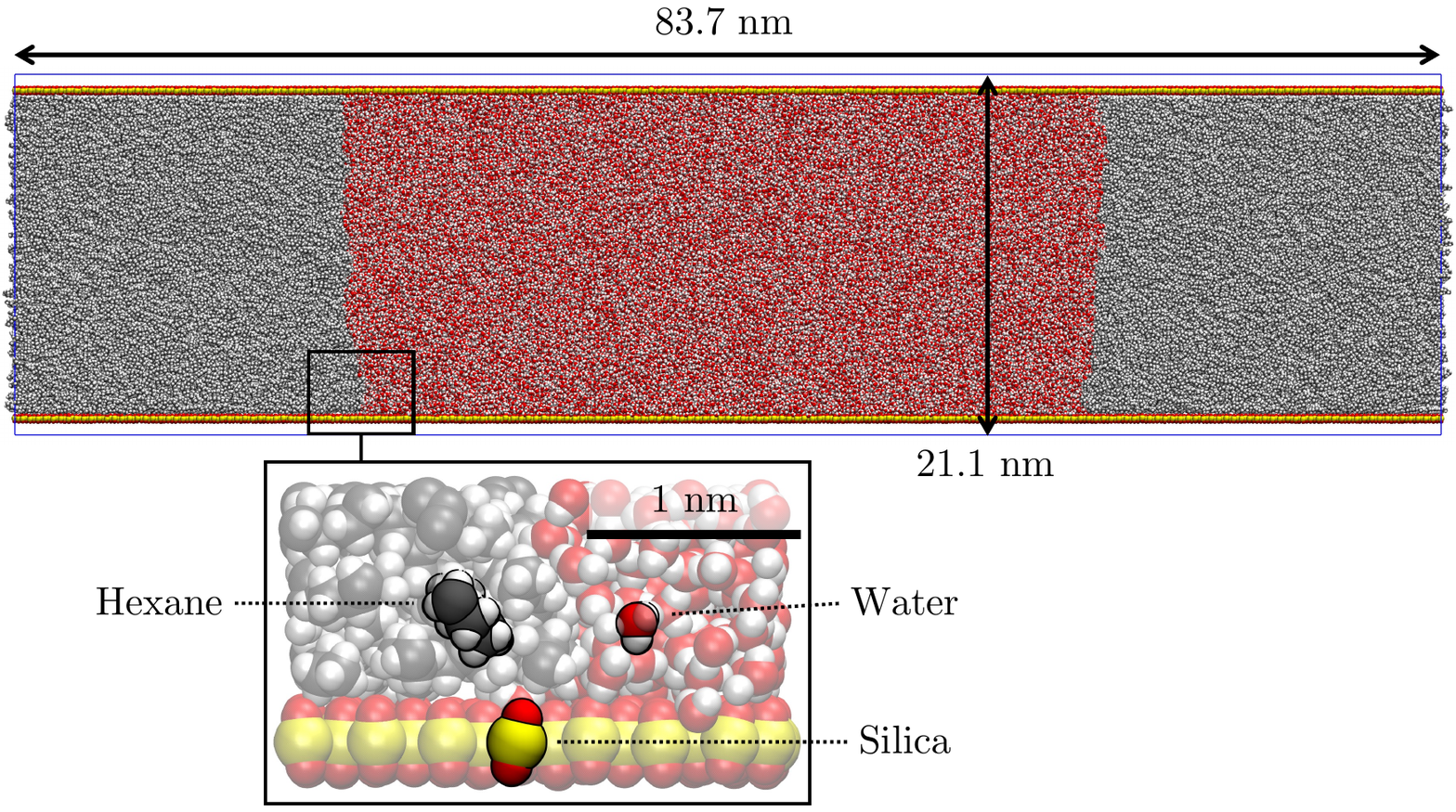}
    \caption{Representation of the molecular system viewed from the $xz$ plane. The length scales reported in the figure indicate the horizontal and vertical size of the periodic box. The inset shows a zoom on the bottom-left contact line, and highlights the 3 molecular species: water, hexane and silica quadrupoles.}
    \label{fig:md-configuration}
\end{figure}
The molecular systems are composed of 128440 water molecules, 15624 hexane molecules and 4464 solid molecules. The systems are simulated in cubic boxes of linear dimensions $(L_x=83.7\mbox{ nm},L_y=4.6\mbox{ nm},L_z=21.1\mbox{ nm})$, with periodic boundary conditions on all edges (fig. \ref{fig:md-configuration}). 

Molecular interactions include short-range van der Waals forces (modelled via the Lennard-Jones potential), long range Coulomb electrostatic forces (treated with the Particle-Mesh-Ewald method), covalent bonds, angles and dihedrals (treated as constraints or listed forces). Water molecules are parametrized by the SPC/E model (Single Point Charge, Extended) by \cite{berendsen1987spce}, while hexane molecules are parameterized via the OPLS-AA force field (Optimised Potentials for Liquid Simulations, All-Atom) by \cite{jorgensen1996oplsaa}. Solid molecules are linear silica quadrupoles (SiO$_2$) arranged in an hexagonal lattice with spacing $d_l=0.45$ nm. Silica molecules are restrained to the lattice by a quadratic spring potential. Partial electric charges are placed on oxygen and silicon atoms in the quadrupole ($q_O=-0.5\cdot q_{Si}=q$); charges can be tuned to achieve different equilibrium contact angles. While this model for the solid substrate is not realistic, in the sense that it does not reproduce the conformation of amorphous or crystalline silica, it has proved effective in reproducing an almost ideal no-slip boundary condition \citep{johansson2015physicochemistry,johansson2019friction}. It is furthermore relatively easy to parametrise. The inset in Figure~\ref{fig:md-configuration} shows a close-up of the near-contact line molecular arrangement. The choice of materials is motivated, on the one hand, by the immiscibility of water and hexane, which can be readily captured through the free-energy definition in the CHNS model, and, on the other hand, by the possibility of forming hydrogen bonds between water and silica, thereby reproducing the physical conditions associated with the absence of slip on hydrophilic surfaces. The 3-point rigid water model was selected by considering the trade-off between computational complexity and the capability to reproduce rheological properties of liquid water.

Interaction parameters for non-bonded forces between different molecular species are generated via geometric combining rules, consistently with the choice of OPLS-AA. All bonds with hydrogen atoms are treated as constraints, while all other bonds and angles are modelled as springs. Atomic positions and velocities are propagated in time using the leapfrog integrator with time-step $\delta t=2$ fs; this is a common choice for simulations that treats C-H and O-H bonds as constraints, and falls well within the stability region of the leapfrog integrator. All simulations are performed with a modified version of GROMACS 2023 \citep{gromacs2015}. For additional technical information on molecular simulations parameters and configuration files, we refer to the GROMACS official documentation (\hyperlink{https://manual.gromacs.org/2023/}{manual.gromacs.org/2023}) and to the configuration files deposited on Zenodo \citep{zenodohydrophobic,zenodohydrophilic}.

\subsection{Simulation setup}

Simulation boxes are prepared by inserting water molecules in the central rectangular region corresponding to 1/2 of the total length of the simulation box and solvating the rest with hexane. The initial vertical coordinate of silicon atoms in the quadrupoles forming the solid walls is placed 1 nm from the box edge. The system is then equilibrated at constant temperature and pressure, respectively 300 K and 1 bar. Temperature control is achieved via the stochastic velocity rescale thermostat by \cite{bussi2007thermostat} (\texttt{v-rescale} in GROMACS), while pressure is controlled with the stochastic cell rescale barostat by \cite{bernetti2020barostat} (\texttt{c-rescale} in GROMACS). The simulation box is allowed to deform only in the vertical direction (i.e. perpendicular to the solid walls). The centre of mass of the reference positions of restrained wall atoms are scaled with the simulation box, effectively allowing the wall location to adapt to the compression/dilation of the liquids without deforming the geometry of the walls. In the equilibration phase both the thermostat and the barostat are applied every 1 ps and last until both temperature and the box dimension are stationary. The same thermostat is applied in the non-equilibrium shear simulations, while the barostat is deactivated. It is important to note that the use of a stochastic velocity rescaling algorithm under non-equilibrium conditions can produce artificial cooling, also known as the ``flying ice cube effect''. This artifact can be mitigated or entirely removed by employing a profile-biased thermostat \citep{bernardi2010thermostat} or a dissipative particle dynamics thermostat \citep{soddemann2003dpd}. In our case, we estimate that the use of velocity rescaling leads at worst to a local temperature deviation of less than 0.1 K. Therefore, we do not consider this effect to be substantial.

To produce a two-phase Couette flow, the lower and upper walls are displaced in opposite directions while the volume and temperature of the system are kept constant. This imposes a flow symmetry whereby the position of the centre of mass of the system is conserved, except for drifts caused by thermal fluctuations and numerical round-off errors. If the resulting flow is steady, the magnitude of the contact line speed $u_{cl}$ in the frame of reference of the moving wall equates the one of the moving walls in the laboratory frame of reference. By convention, the speed of advancing contact lines has positive sign and the one of receding contact lines has negative sign. The non-equilibrium simulations are performed for $u_w\in[1.12,4.84]$ m/s and for a total time ranging between $\sim$50 and $\sim$90 ns, depending on the wall speed and the surface wettability.

Due to the sheer size of the molecular system and the length of the simulations, it is impractical to output, store and analyse frequently-sampled trajectories containing atomic positions and coordinates. Moreover, the quasi-2D nature of the system can be exploited both for the extraction of the liquid/liquid interface location and for the comparison with 2D CHNS equations. Density, velocity and temperature fields for both liquids are binned \textit{on-the-fly} on a 2D homogeneous structured grid in the $xz$ plane, averaging over the direction spanned by $y$. The grid contains $N_x=419$ cells along $x$ and $N_z=105$ cells along $z$. Flow fields are also averaged in time in 20 ps windows, and stored after averaging. Further details regarding system preparation, equilibration and non-equilibrium runs can be found in the supplementary information.

\subsection{Observables}

\begin{figure}
    \centering
    \includegraphics[width=0.6\linewidth]{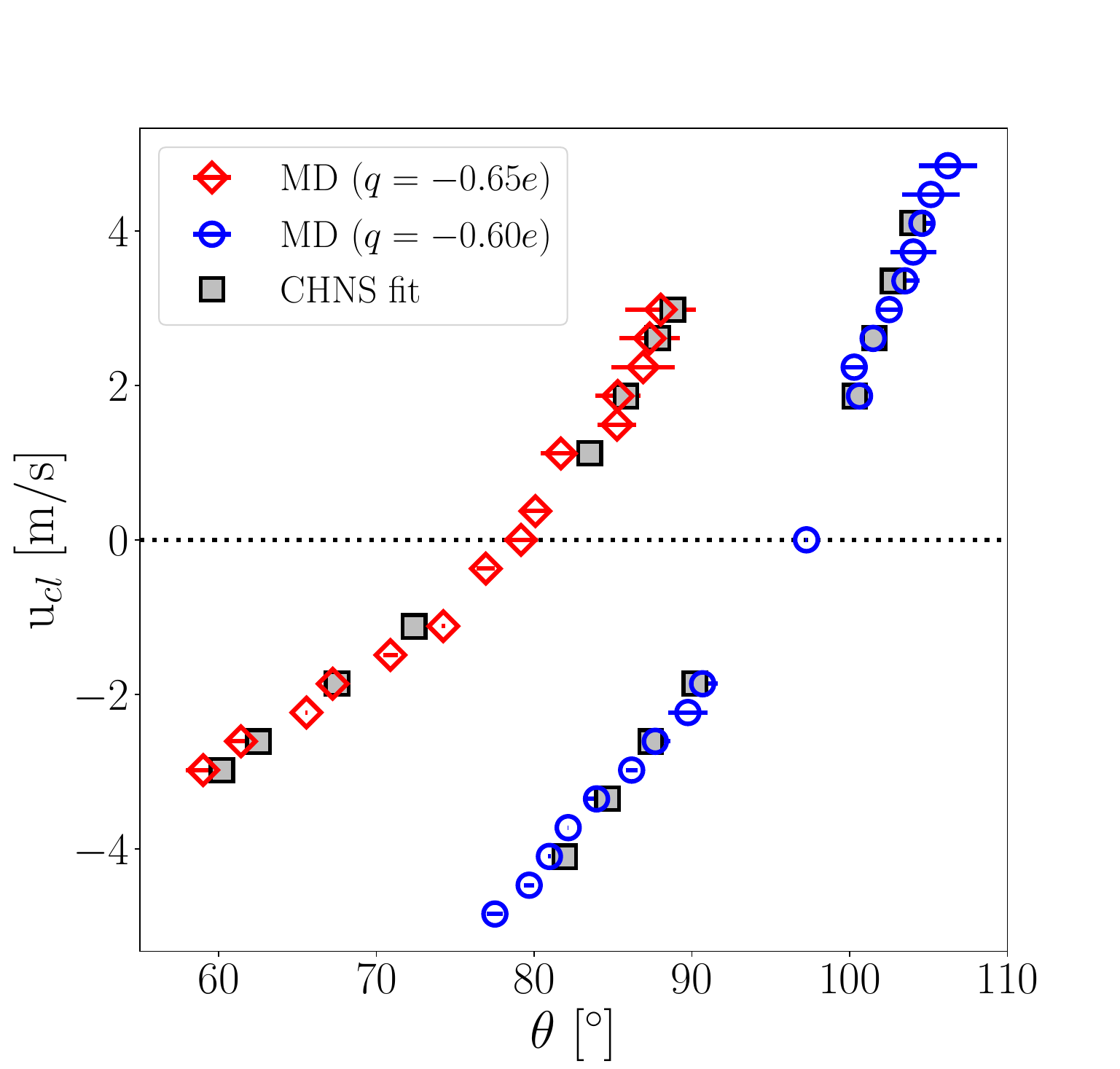}
    \caption{Relation between contact line speed and dynamic contact angle for $\theta_0\simeq97.3^{\circ}$ (blue circles) and $\theta_0\simeq80.9^{\circ}$ (red diamonds), determined from MD simulations. Error bars refer to the uncertainty on the steady dynamic contact angle. Square markers indicate the dynamic contact angles computed from CHNS simulations, obtained through the calibration procedure discussed in Section~\ref{sec:md-vs-pf}.}
    \label{fig:md-observables}
\end{figure}

The detection of the liquid/liquid interface and the measurement of contact line friction follow the same procedure of \cite{lacis2022pellegrino}. Details on the interface extraction are presented in appendix \ref{sec:app-md}. Figure~\ref{fig:md-observables} shows the dynamic contact angle as a function of the speed of the contact lines. As it can be noted by observing the different trend for negative and positive contact line speed, advancing and receding contact lines experience different friction. This phenomenon has been previously observed for water/vapour interfaces moving on the same type of surface \citep{pellegrino2022asymmetry}. We will therefore fit two separate contact line friction coefficients for advancing and receding contact lines. It is worth remarking that the different contact line friction on advancing and receding sides is related to non-equilibrium dynamic behaviour, and should be distinguished from \textit{contact angle hysteresis}, which refers to differences between advancing and receding contact angles at the onset of contact line motion. The shear viscosity coefficients of water and hexane can also be measured from molecular simulations; they are respectively $\eta_w=0.69$ cP and $\eta_h=0.36$ cP (we refer to appendix \ref{sec:app-md} for the details on the calculation).

\section{Comparison between simulations techniques}
\label{sec:md-vs-pf}


\begin{table}
\begin{center}
\begin{tabular}{p{12mm} | p{24mm} p{24mm} p{12mm} p{28mm} p{14mm} }
 $\theta_0$ [$^\circ$] & $\mu_f^{adv}$ [cP] & $\mu_f^{rec}$ [cP] & $\varepsilon$ [nm] & $M$ [nm$^2$ps$^{-1}$bar$^{-1}$] & $\ell_N$ [nm] \\
  & & & & & \\
 97.3 & $2.86$ & $7.09$ & 0.3 & $9.04\cdot10^{-6}$ & $5\cdot10^{-3}$ \\
 80.9 & $4.71$ & $14.09$ & 0.3 & $9.04\cdot10^{-6}$ & $5\cdot10^{-4}$ \\
\end{tabular}
\end{center}
\caption{Set of parameters obtained after the calibration procedure. Equilibrium contact angles are computed directly from MD simulations, while all other parameters are calibrated under the constraint of Equation~\ref{eq:sharp-interface-limit} to reproduce MD results. The $95\%$ intervals on $\mu_f$ carry the same relative tolerance as the MD friction coefficients $\mu_f^{*}$ (Table~\ref{tab:sharp-interface-slip-md}).}
\label{tab:cl-steady-param-match}
\end{table}

We calibrate the CHNS model following a systematic procedure. We first select the value of the interface thickness $\varepsilon$ that matches the MD simulation results (see Appendix~\ref{sec:app-md}). Contact line friction coefficients ($\mu_f^{\text{adv}}$ and $\mu_f^{\text{rec}}$) are then extracted empirically from MD contact angle data (Figure~\ref{fig:md-observables}), fixing the threshold mobility $M$ from the sharp interface condition. The equilibrium and dynamic contact angles are shown to be independent of $M$: this allows us to capture the correct interface curvature, as a first essential step. We subsequently refine $M$, still constrained by Equation~\ref{eq:sharp-interface-limit}, to match the steady contact line displacement observed in MD simulations. We then verify the flow field to ensure there is minimal streamline crossing, confirming $M$ is in the appropriate range. The Navier slip length $\ell_N$ is not treated as a fitting parameter: it is fixed at a sub-nanometric value (of order $10^{-3}$~nm, well below atomic length scales and therefore not physically meaningful) solely to ensure numerical stability, and we verify in Section~\ref{sec:sensitivity} that it has a weak effect on the steady contact line displacement for values in the range considered. This multi-observable approach ensures that the model accurately captures contact line physics, while simultaneously testing its robustness. The values of the parameters resulting from the calibration procedure are listed in Table~\ref{tab:cl-steady-param-match}.

\subsection{Interfacial curvature}    \label{sec:curvature}

\begin{figure}
    \centering
    \includegraphics[width=\linewidth,trim={0 2cm 0 2cm},clip]{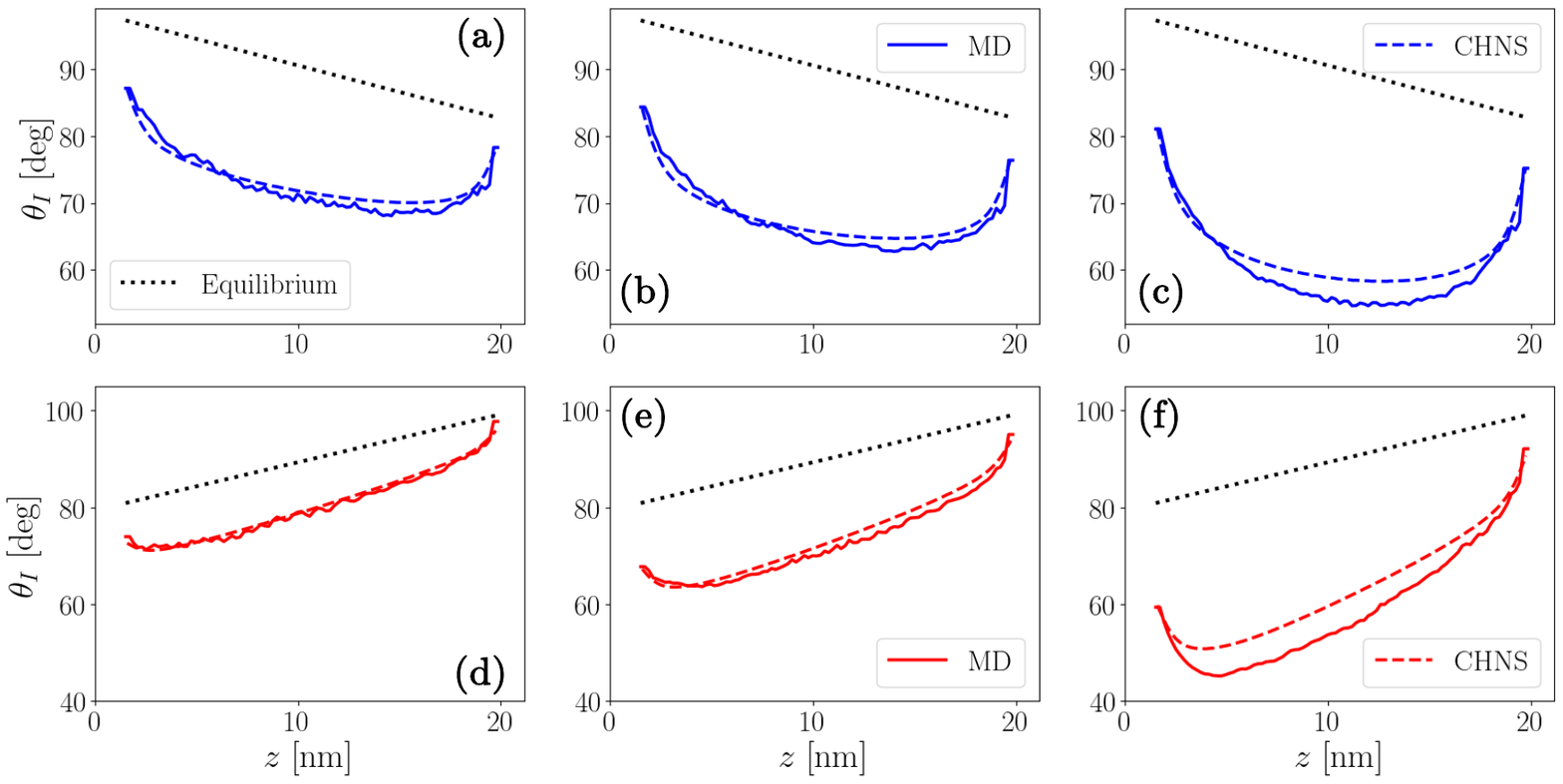}
    \caption{Interface curvature profiles for hydrophobic surfaces (top: $u_{cl}=$ (a) 2.61 m/s, (b) 3.35 m/s, (c) 4.1 m/s) and hydrophilic surfaces (bottom: $u_{cl}=$ (d) 1.12 m/s, (e) 1.86 m/s, (f) 2.98 m/s). The interface angle $\theta_I$ is computed taking the lower wall as reference. Therefore, the contact angle in the frame of reference of the upper wall would be $\pi-\theta_I$.}
    \label{fig:compare-int-curvature}
\end{figure}

The values of $\mu_f^{adv}$ and $\mu_f^{rec}$ are tuned to reconstruct the profiles of the contact line speed against the dynamic contact angle (Figure~\ref{fig:md-observables}). This choice stems from the interpenetration given by Molecular Kinetic theory to the contact line friction coefficient, which measures the response of the uncompensated Young stress (or alternatively the contact angle) to the motion of the contact line.

To compare the interfacial curvature, we compute the angle $\theta_I(z)$ between the steady liquid/liquid interface and the $x$-parallel plane (see Figure \ref{fig:chns-geometry}). Results are shown in Figure~\ref{fig:compare-int-curvature}. The reconfiguration of liquid/liquid interfaces is an effect of the two main channels of energy dissipation: contact line friction and viscosity. Contact line friction produces a difference between equilibrium and dynamic contact angle, and hence is responsible for the shifting the profile of $\theta_I(z)$ downwards w.r.t. to the straight line corresponding to the equilibrium interface. On the other hand, viscosity produces a deviation of the interface curvature from equilibrium, i.e. the inflection of $\theta_I(z)$. As we will discuss in Section~\ref{sec:sensitivity}, the dynamic contact angle is not affected by Cahn-Hilliard mobility $M$, while the interface curvature profile is.

The profiles of $\theta_I(z)$ show quantitative agreement in all cases where contact line speed is sufficiently small: this indicates that the Phase Field model is capable of capturing the effect of both contact line friction and viscosity for moderate deviations from equilibrium. The capability of replicating viscous bending is not surprising: this is essentially an outcome of Navier-Stokes equations, and thus it should be captured by continuous hydrodynamics. The accentuation of interface curvature upon increasing contact line speed agrees with the results by \citet{fullana2025}, which derived a scaling in the absence of contact line friction. On the other hand, the reconstruction of interface curvature in the \textit{near-wall} region is mainly due to contact line friction. This represents an advancement with respect to previous studies, as it clearly shows that microscopic contact angle relaxation is as important as viscous bending in determining \textit{apparent} contact angles, corresponding to zero-slope points in the graphs of figure \ref{fig:compare-int-curvature}.

\subsection{Steady contact line displacement}

After a sufficiently long transient, and for sufficiently small wall velocities, a steady two-phase Couette flow develops. The steady displacement between the advancing and the receding contact lines $\Delta x_{cl}$ depends on the wall speed and it is found to be sensitive to slip parameters $M$ and $\ell_N$ (see Section~\ref{sec:sensitivity}). Hence, it is reasonable to choose it as reference observable to calibrate CHNS equations against. The calibration is constrained by the SIL Equation~\ref{eq:sharp-interface-limit}, so that the values of $M$ and $\varepsilon$ are bounded from below, while $M$ and $\ell_N$ are varied to minimise the difference with respect to the time-averaged steady displacement obtained from MD simulations.
\begin{figure}
    \centering
    \includegraphics[trim={0 2cm 0 2cm},clip,width=0.85\linewidth]{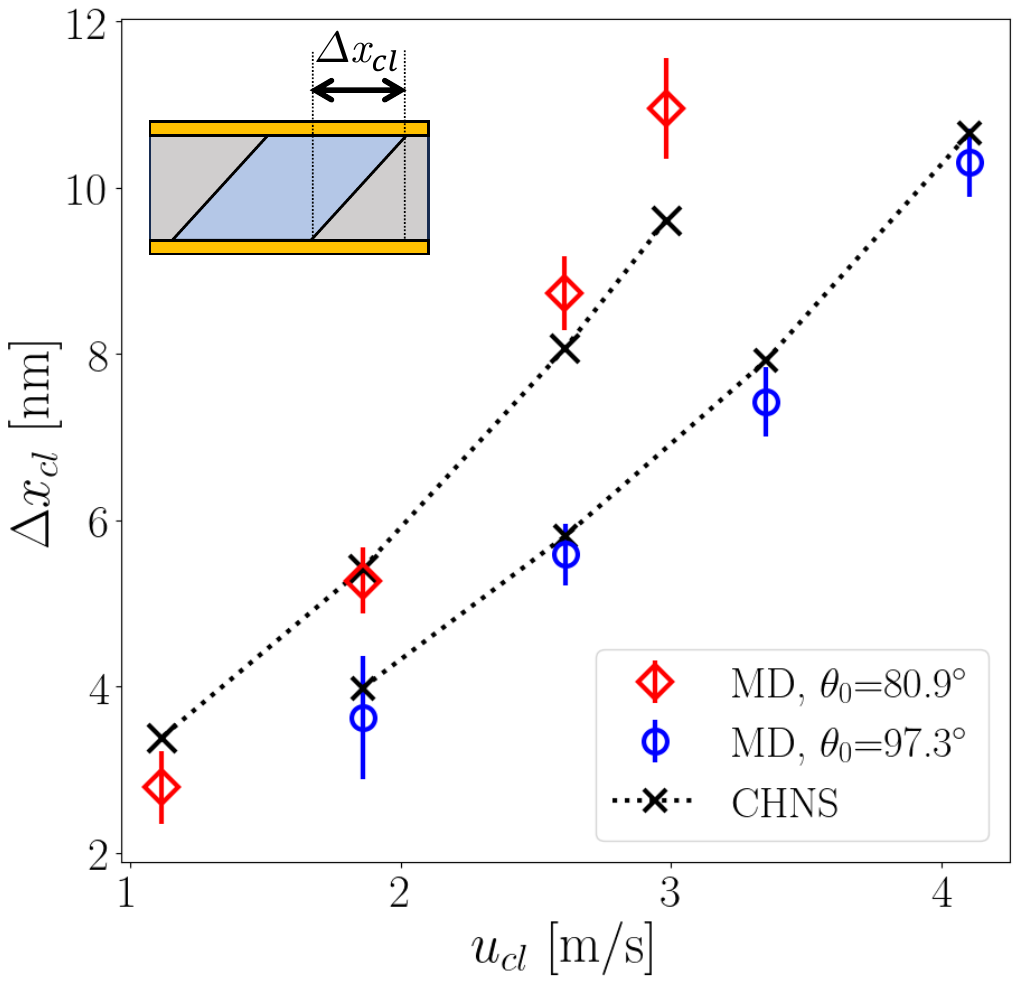}
    \caption{Steady displacement against contact line speed. The error bars show the standard deviation of the contact line displacement computed from MD simulations.}
    \label{fig:compare-steady-disp}
\end{figure}

The results shown in Figure~\ref{fig:compare-steady-disp} correspond to the choice of parameters reported in Table~\ref{tab:cl-steady-param-match}. For the hydrophobic substrate ($\theta_0 = 97.3^\circ$), the calibration is satisfactory: all results from CHNS simulations fall within the uncertainty of the average displacement computed from MD simulations across the full range of contact line speeds investigated. For the hydrophilic substrate ($\theta_0 = 80.9^\circ$), the agreement is similarly good at lower contact line speeds, but the Phase Field simulations systematically underestimate the steady displacement at higher speeds. This growing discrepancy with increasing velocity can be attributed to the absence of the diffusive flux coupling term in the Navier--Stokes momentum equation, present in the thermodynamically consistent formulations of \cite{abels2012consistent}, \cite{yue2020thermodynamically} or \cite{brunk2026chns}. This term becomes increasingly important at higher contact line speeds, where the diffusive mass flux across the interface contributes significantly to momentum transport. We note however that the water/hexane system considered here has a comparatively moderate density contrast ($\rho_w/\rho_h \simeq 1.5$); since this coupling term arises from, and scales with, the density difference between the two phases, its contribution in the present configuration is correspondingly limited and becomes appreciable only at the largest contact line speeds investigated.

We remark that the same values of $M$, $\varepsilon$, and $\ell_N$ are used for both substrates and across the entire range of wetting speeds. Only the contact line friction coefficients ($\mu_f^{adv}$ and $\mu_f^{rec}$) differ between the hydrophilic and hydrophobic surfaces, as these are extracted empirically from the MD results. This separation can be physically motivated; for example, $M$ influences the steady-state displacement through its contribution to diffusive slip at the contact line (see Section~\ref{sec:sensitivity}), but does not affect the dynamic contact angle since $M$ does not appear in the wall boundary condition (Equation~\ref{eq:wet-bc}). In contrast, the contact line friction coefficient $\mu_f$ directly governs the relaxation of the contact angle towards equilibrium, and is therefore the sole parameter responsible for capturing the correct dynamic contact angle behaviour.

The only arbitrary parameter obtained via indirect calibration, the Navier slip length $\ell_N$, is sub-nanometric, in accordance with the physics of hydrogen-bond mediated wetting. While the Navier slip length is very small, it is not zero. We identify two needs for introducing a non-zero slip length. On one hand, the sharp interface limit is attained only numerically; the thickness of interface anti-correlates with the steady contact line displacement, i.e.\ a wider interface is characterised by a lower diffusive slip. Hence, some non-zero Navier slip may be introduced to compensate for the rigidity of the interface. On the other hand, Navier slip may be introduced for the purpose of ensuring numerical stability. Stable simulations can be obtained also without introducing Navier slip, but require a smaller integration time step.

\subsection{Flow field}

\begin{figure}
    \centering
    \includegraphics[width=\linewidth,trim={0 2cm 0 2cm},clip]{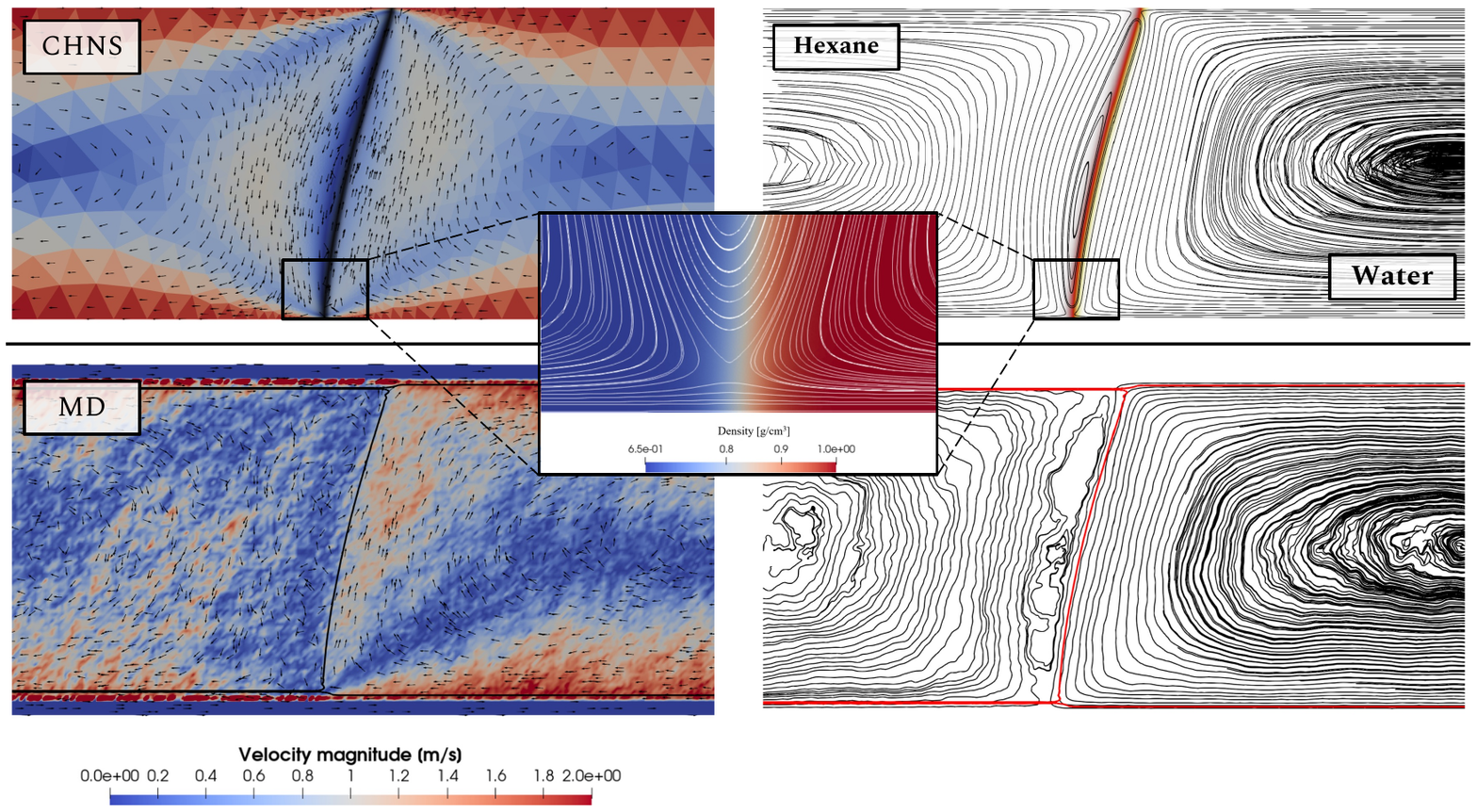}
    \caption{Comparison of the velocity field and the streamline profiles of CHNS simulations and MD simulations for $u_w=1.86$ m/s and $\theta_0=97.3^{\circ}$. The black lines on the left panels and the red lines in the right panels indicate the liquid/liquid interfaces. Inset: streamline profile in the contact line region.}
    \label{fig:compare-streamlines}
\end{figure}

The flow fields are then compared both qualitatively to determine the effect of $M$ on the streamlines and interface reconfiguration. The flow of the lower-viscosity liquid, hexane in our case, has to `adapt' to the flow of the higher-viscosity liquid, water in our case, given that the tangential component of flow velocity to the interface remains continuous. This provokes the formation of a recirculation region and streamlines \textit{splitting} close to the contact line in the hexane phase, which is captured by CHNS as Figure~\ref{fig:compare-streamlines} shows. The only quantitative discrepancy between Molecular Dynamics and Phase Field simulations lies in the size of the eddy in the hexane phase near the liquid/liquid interface. It should be noted that the maximum width of the eddy in Phase Field simulations is comparable to the length of two hexane chains: this type of sub-continuous flow cannot be accurately described by a continuous model that doesn't account for non-local steric interactions between molecules.

We observe that there is no streamline crossing the liquid/liquid interface in the CHNS simulations, which match the MD simulations well in that aspect. Streamline crossing is known to become stronger as $M$ is increased \citep{lacis2022pellegrino}, and this suggests the value of $M$ chosen according to the sharp interface condition is in the right range. We also observe the position of the recirculating region, and hence the stagnation point to migrate towards the hexane phase. This observation are in agreement with the ones by \cite{yue2010sil}. Overall, the velocity fields calculated from MD and CHNS simulations, as shown in Figure~\ref{fig:compare-streamlines} agree qualitatively with the recirculating region in the hexane phase having higher velocity compared to the region close to the interface on the water phase. But we also notice slight differences in the bulk phase of hexane in both MD and CHNS simulations than when compared with that of water.

\section{Discussion}
\label{sec:discussion}

\subsection{Sensitivity analysis}   \label{sec:sensitivity}

Having established the set of parameters reported in Table~\ref{tab:cl-steady-param-match} through the calibration procedure, we now examine the sensitivity of the results to variations in these parameters. This serves two purposes: to assess the robustness of the calibration, and to clarify the distinct physical role of each parameter in controlling the observables of interest.

\begin{figure}
    \centering
    \includegraphics[trim={0 2cm 0 2cm},clip,width=\textwidth]{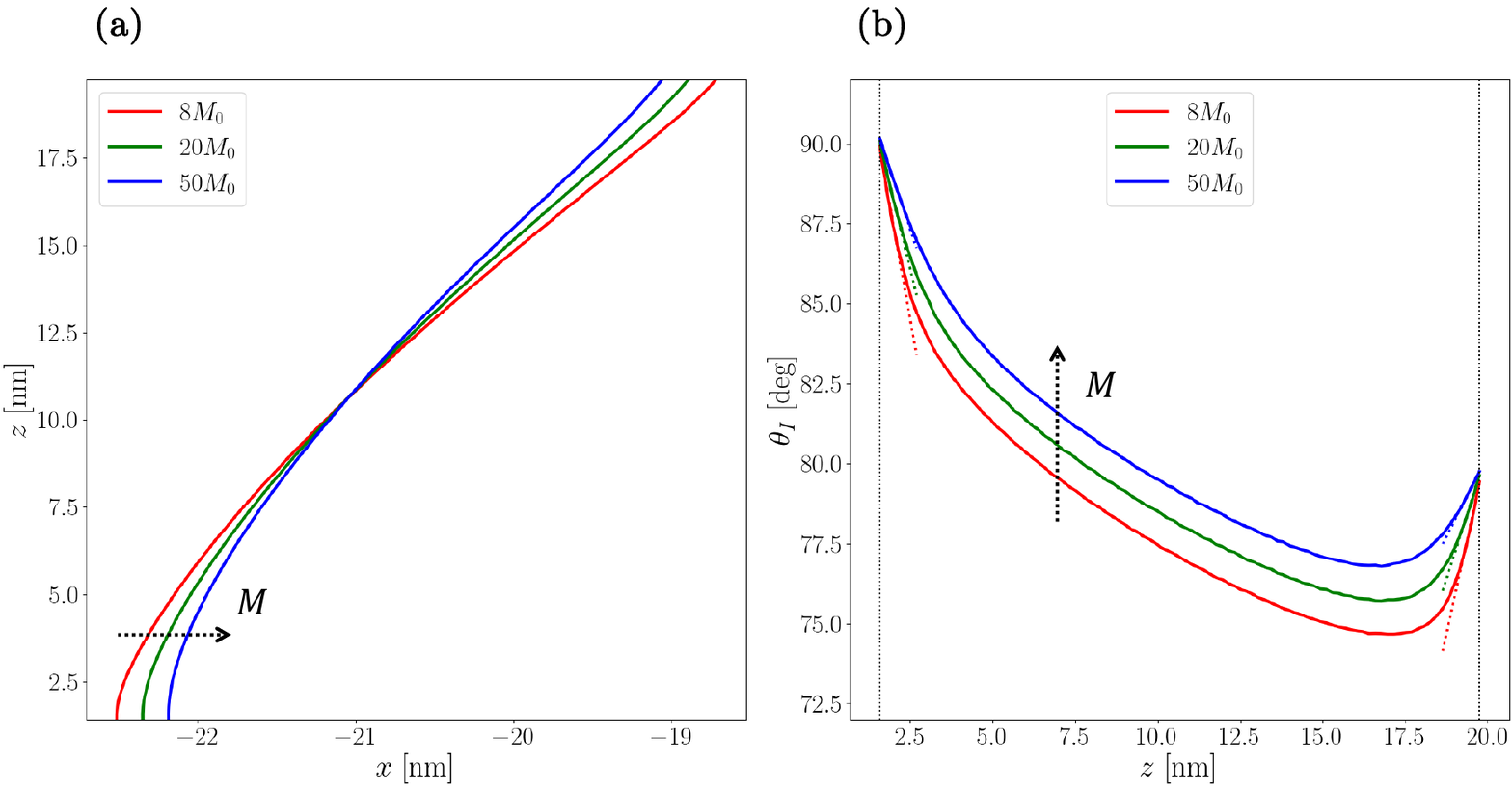}
    \caption{Effect of increasing the mobility parameter to the interface profile (a) and interface bending (b), $\theta_0=97.3^{\circ}$, $u_{cl}=1.86$ m/s. The reference value is the lower bound of Equation~\ref{eq:sharp-interface-limit}, that is: $M_0=\varepsilon^2/(16\eta^*)$. Coloured dotted lines highlight the variation of near-wall interface curvature upon changing $M$.}
    \label{fig:sensitivity-mobility}
\end{figure}

We begin by examining the role of $M$. The mobility parameter affects the steady contact line displacement $\Delta x_{cl}$ through its contribution to diffusive slip at the contact line as shown in Figure~\ref{fig:sensitivity-mobility}a, which adds constructively with $\ell_N$, and it is therefore calibrated against $\Delta x_{cl}$ under the constraint of Equation \ref{eq:sharp-interface-limit}. Varying $M$ within the range permitted by the sharp interface condition (Equation~\ref{eq:sharp-interface-limit}) leaves the dynamic contact angle $\theta$ unchanged: this can clearly be observed in Figure~\ref{fig:sensitivity-mobility}b, where the value of $\theta_I$ at the walls remains unchanged. However, $M$ does influence the interface curvature profile away from the wall (as shown again in Figure~\ref{fig:sensitivity-mobility}b), with larger $M$ producing less pronounced interfacial bending. This observation is consistent with the results by \citet{yue2011wallrelax}, showing a similar effect of $M$ on the apparent macroscopic contact angle.

\begin{table}
\begin{center}
\begin{minipage}[t]{0.24\textwidth}
\vspace{15pt}
\includegraphics[width=\textwidth]{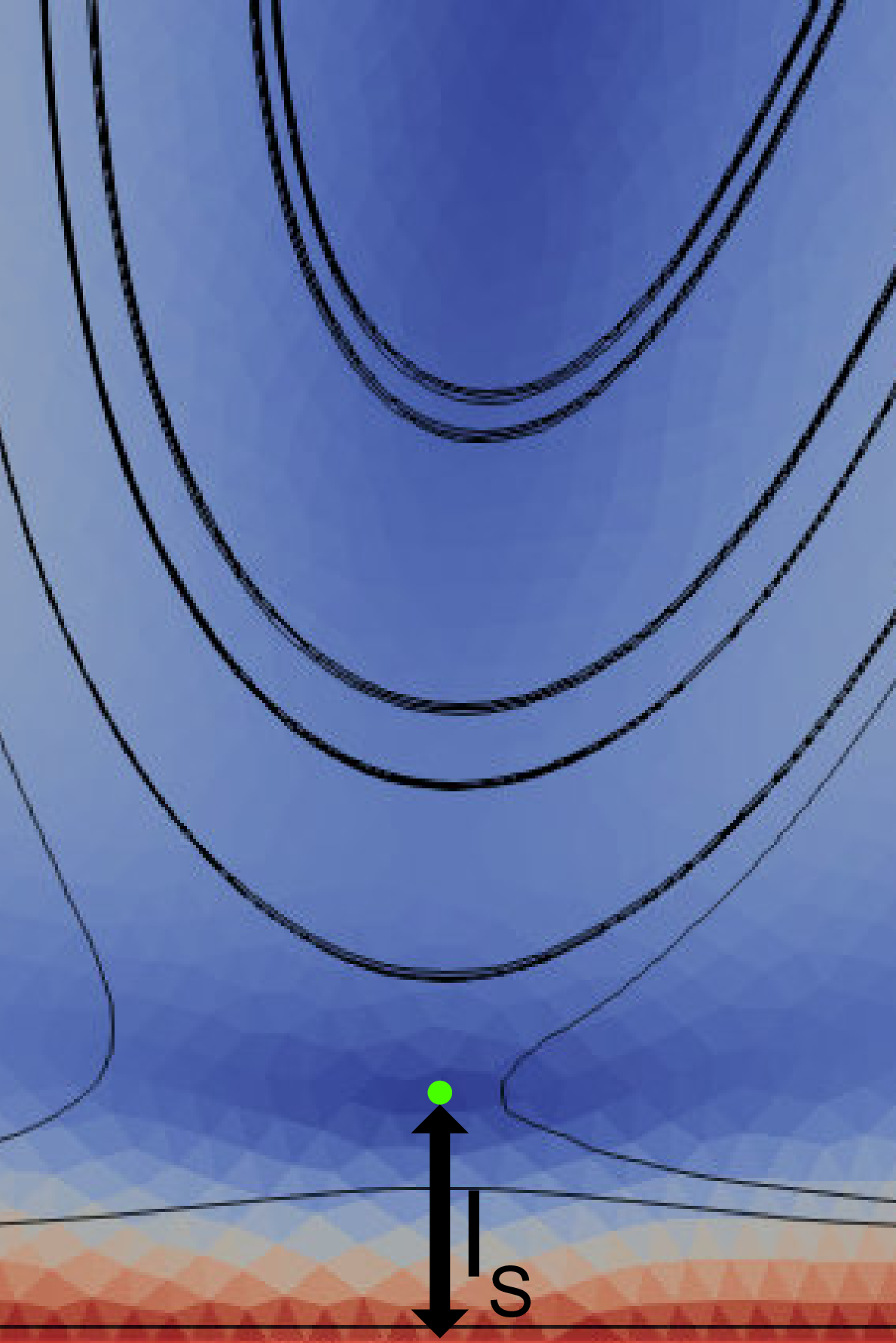}
\end{minipage}
\hfill
\begin{minipage}[t]{0.74\textwidth}
\vspace{0pt}
\begin{tabular}{p{10mm} p{25mm} p{12mm} | p{12mm} p{12mm} p{12mm} }
Ca & $M$ & $\ell_N$ & $\ell_S$ & $\ell_D$ &$\ell_S/\ell_D$ \\ 
& [nm$^2$ps$^{-1}$bar$^{-1}$] & [nm] & [nm] &  [nm] & \\
 & & & & & \\
0.05 & 1.13$\cdot10^{-6}$ & 5$\cdot10^{-3}$ & 0.22 & 0.075 & 2.93 \\
0.07 & 1.13$\cdot10^{-6}$ & 5$\cdot10^{-3}$ & 0.22 & 0.075 & 2.93 \\
0.09 & 1.13$\cdot10^{-6}$ & 5$\cdot10^{-3}$ & 0.22 & 0.075 & 2.93 \\
0.11 & 1.13$\cdot10^{-6}$ & 5$\cdot10^{-3}$ & 0.22 & 0.075 & 2.93 \\ \hline
0.05 & 1.13$\cdot10^{-6}$ & 5$\cdot10^{-3}$ & 0.22 & 0.075 & 2.93 \\
0.05 & 5.65$\cdot10^{-6}$ & 5$\cdot10^{-3}$ & 0.4220& 0.1678 &2.52 \\
0.05 & 11.3$\cdot10^{-6}$ & 5$\cdot10^{-3}$ & 0.6055& 0.2373 &2.55 \\
0.05 & 22.6$\cdot10^{-6}$ & 5$\cdot10^{-3}$ & 0.8441& 0.3356 &2.52 \\ \hline
0.05 & 1.13$\cdot10^{-6}$ & 5$\cdot10^{-4}$ & 0.22 & 0.075 &2.93 \\
0.05 & 1.13$\cdot10^{-6}$ & 1$\cdot10^{-3}$ & 0.22 & 0.075 &2.93 \\
\end{tabular}
\end{minipage}
\end{center}
\caption{Relation between diffusion length $\ell_D$ and slip length $\ell_S$, obtained by varying one parameter at the time (Ca, $M$ and $\ell_N$). Inset: coordinate of the stagnation point.}
\label{tab:ls-ld}
\end{table}

We also proceed to verify the relation between $\ell_S$, which is the distance from the stagnation point to the wall, or in other words the slip of the contact line due to the flow between the stagnation point and wall, and $\ell_D$, the diffusion length defined as $\ell_D = \sqrt{M\eta^*}$. We vary $M$, $\ell_N$ and $u_w$ and find the relation to be $\ell_S=2.9\ell_D$ for the lowest $M$ possible according to sharp interface condition
and for all $u_w$ and $\ell_N$. Table \ref{tab:ls-ld} gives a detailed view of the relations obtained for various parameters. Here capillary number is defined as $\mbox{Ca}=u_{cl}\eta^*/\sigma$. As we increase $M$ from the strictly minimum permissible value according to inequality in Equation~\ref{eq:sharp-interface-limit} to a conservative value i.e. up to twenty times its minimum value, the constant reduces from $2.9$ and maintains a steady value of around $2.5$. From \cite{yue2010sil}, who had a similar setup, we know this constant to be $2.5$, thereby reinforcing the results we have obtained.

The recent model by \cite{fullana2025} allows to connect the increase of $\ell_D$ with $M$ shown in Table~\ref{tab:ls-ld} with that in Figure~\ref{fig:sensitivity-mobility}b: the radius of curvature of the interface in the contact line region is expected to scale with $\ell_D$, or, in other words, interface bending occurs more smoothly and on a longer distance as $M$ increases. This ``curvature smearing'' effect is expected to be more pronounced in macroscopic flows, where $M$ and $\varepsilon$ are orders of magnitude larger.

\begin{table}
\begin{center}

\begin{tabular}{p{30mm}  p{30mm} | p{30mm} }
\hspace{10mm}{$\Delta H/H$ [$\%$]}           & & $\Delta \Delta x / \Delta x$ [$\%$]     \\
& & \phantom{x} \\
\hspace{14mm}{-1.08}                 & & -0.19                  \\
\hspace{14mm}{1.08}                  & & 0.39   \\
\end{tabular}

\vspace{3pt}
\noindent\rule{\dimexpr 90mm + 6\tabcolsep\relax}{0.4pt}
\vspace{3pt}

\begin{tabular}{p{30mm}  p{30mm} | p{30mm} }
\hspace{10mm}{$\Delta \ell_N/\ell_N$ [$\%$]}     & & $\Delta \Delta x / \Delta x$ [$\%$] \\
& & \phantom{x} \\
\hspace{14mm}{400}                    & & -1.01               \\
\hspace{14mm}{2400}                   & & -5.4              \\ 
\end{tabular}

\end{center}
\caption{Sensitivity of  $H$ in CHNS simulations, whose relative perturbation corresponds to the uncertainty on the wall location. Values for $\ell_N$ sensitivity analysis are logarithmically spaced.}
\label{tab:muf-sen}
\end{table}

Lastly, in order to understand the degree of effect the error in $H$ and $\ell_N$ have on the water-hexane system, we perturb the values chosen for the simulations. Table \ref{tab:muf-sen} shows the results of this sensitivity analysis. We perturb $H$ by $\pm 0.2$ nm, corresponding to the average grid size used to post-process the MD simulations. We see that there is negligible change in the steady state displacement, $\Delta x$, of the water-hexane system, confirming that the uncertainty in $H$ has negligible effect on the results. Regarding $\ell_N$, it is a free parameter introduced to circumvent the numerical instability arising from a zero slip length. We vary the $\ell_N$ chosen in simulations ie. $\ell_N=0.005$ nm by a reasonable amount to see the difference in steady state displacement with respect to increase in $\ell_N$ as shown by $\Delta \ell_N$, and conclude that $\Delta x$ is insensitive to $\ell_N$ within the range tested.

\subsection{Sharp interface limit with slip}    \label{sec:sil-slip-md}

Attaining the sharp interface limit means producing a physically meaningful, converged and unique solution by taking $\varepsilon\rightarrow0$, $M\rightarrow0$, under the constraint of inequality \ref{eq:sharp-interface-limit}. Practically, a \textit{numerical} sharp interface limit can be achieved by testing a set of increasingly smaller values of $\varepsilon$ and $M$ until the solution computed numerically becomes insensitive to these two parameters. For sufficiently small contact line velocities and in the absence of Navier slip at the liquid/solid interface, \cite{yue2011wallrelax} showed that the wetting boundary condition \ref{eq:wet-bc} reduces to:
\begin{equation}    \label{eq:cl-friciton-sharp-interface-limit}
    \Big(\frac{2\sqrt{2}}{3}\sin\theta\Big)\mu_f^* u_{cl} = \sigma\big(\cos\theta_0-\cos\theta\big) \; .
\end{equation}
Equation~\ref{eq:cl-friciton-sharp-interface-limit} can be regarded as a local \textit{mobility relation} for the contact line, equating the energy dissipated by contact line friction to the work due to the uncompensated Young stress, analogous to the one derived by Molecular Kinetic Theory. This expression is also similar to that proposed by \cite{fullana2026gnbc} for the Contact Region Generalized Navier Boundary Condition (CR-GNBC), obtained by regularising the uncompensated Young stress over a region of finite width under a quasi-stationary reduction; the only formal difference between the two expressions is the factor $2\sqrt{2}(\sin\theta)/3$ at the LHS of \ref{eq:cl-friciton-sharp-interface-limit}, which is close to unity for moderately hydrophobic or hydrophilic surfaces.

\begin{figure}
    \centering
    \includegraphics[trim={0 2.75cm 0 2.75cm},clip,width=\textwidth]{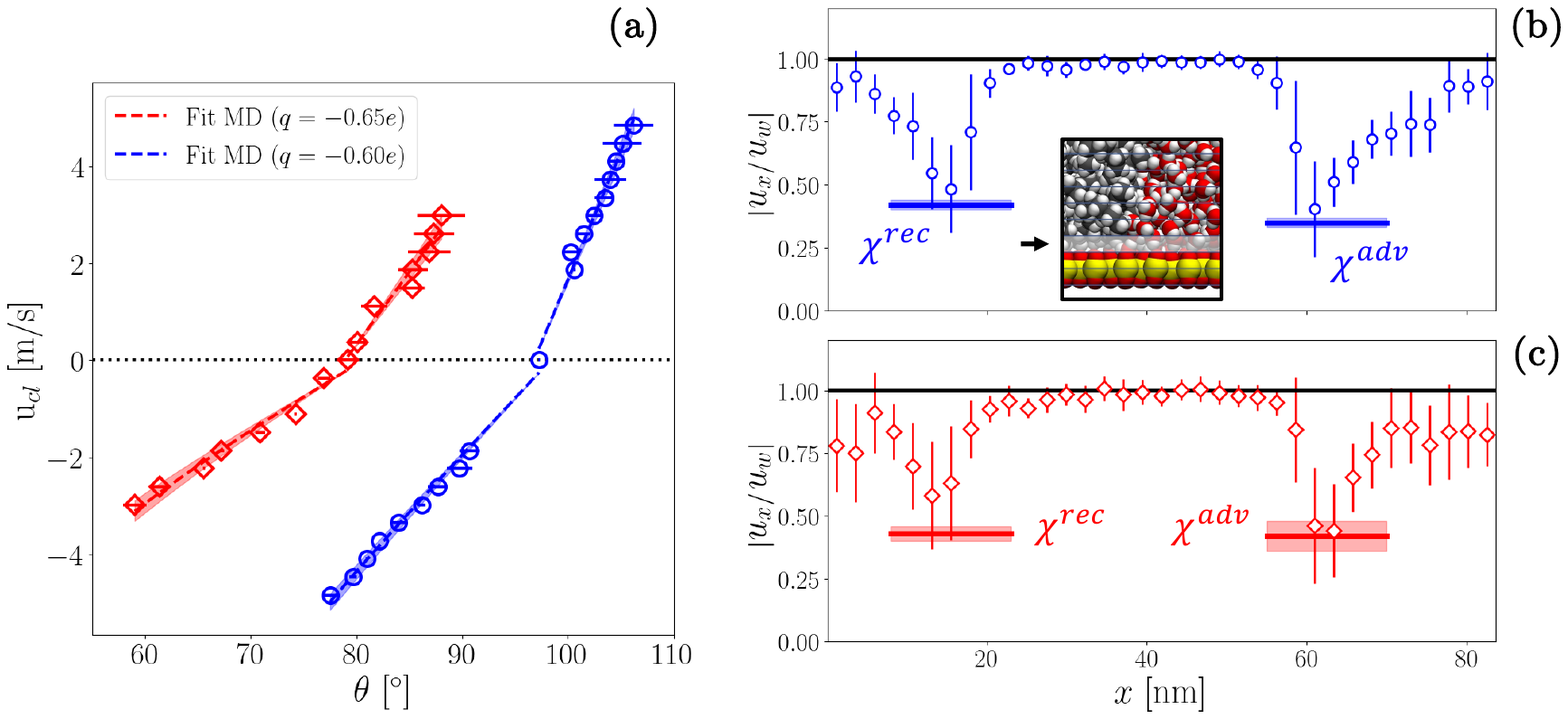}
    \caption{Left: dynamic contact angle measured from MD simulations and fit of Equation~\ref{eq:cl-friciton-sharp-interface-limit}. Right: horizontal velocity profile scaled by the wall velocity in the first density bin above the wall (inset), respectively for the case $\theta_0=97.3^{\circ}$ (b, $\circ$) and the case $\theta_0=80.9^{\circ}$ (c, $\diamond$). Data are obtained by collecting and collapsing the velocity fields of all simulations, i.e. for all imposed wall velocities; error bars indicate the standard deviation between simulations. The values of $\chi$ are reported as straight lines ($\pm$ error), in correspondence of the location of the advancing and receding contact line regions.}
    \label{fig:md-slip-velocity}
\end{figure}

\begin{table}
\begin{center}
\begin{tabular}{p{9mm} | p{17mm} p{17mm} | p{17mm} p{15mm} p{15mm} p{15mm} p{15mm} }
 $\theta_0$ [$^\circ$] & $\mu_f^{*,adv}$ [cP] & $\mu_f^{*,rec}$ [cP] & $\zeta$ & $\chi^{adv}$ & $\chi^{rec}$ & $\varepsilon\beta_r^{adv}/\eta^*$ & $\varepsilon\beta_r^{rec}/\eta^*$ \\
  & & & & & & \\
 97.3 & 1.86$\pm$0.09 & 4.09$\pm$0.15 & 1.36$\times10^{-4}$ & 0.35$\pm$0.03 & 0.42$\pm$0.02 & 1.68 & 6.33 \\
 80.9 & 2.71$\pm$0.30 & 8.09$\pm$0.43 & 1.36$\times10^{-5}$ & 0.42$\pm$0.06 & 0.43$\pm$0.03 & 4.22 & 13.10 \\
\end{tabular}
\end{center}
\caption{Values of $\mu_f^*$, $\zeta$ and $\chi$ obtained by fitting Equation~\ref{eq:cl-friciton-sharp-interface-limit} and evaluating expressions \ref{eq:expression-zeta} and \ref{eq:expression-chi}. Note that we distinguish the values of $\mu_f^*$ and $\chi$ between advancing and receding contact lines, while $\zeta$ ultimately depend only on the Navier slip length and the geometry of the system. The values of $\beta_r$ are scaled by a factor $\eta^*/\varepsilon$, simply to make them non-dimensional.}
\label{tab:sharp-interface-slip-md}
\end{table}

Figure~\ref{fig:md-slip-velocity}a shows the best fit of Equation~\ref{eq:cl-friciton-sharp-interface-limit} to MD simulation data (wall velocities against measured dynamic contact angles); Table~\ref{tab:sharp-interface-slip-md} reports the values of $\mu_f^*$. By comparing these results with the ones of Table~\ref{tab:cl-steady-param-match}, it is clearly noticeable that the values of $\mu_f^*$ substantially underestimate the ones of $\mu_f$, i.e. the contact line friction parameter obtained by calibrating CHNS. We therefore investigate the source of this discrepancy. Let us first propose a correction of \ref{eq:cl-friciton-sharp-interface-limit} in case of non-zero slip velocity:
\begin{equation}    \label{eq:sil-correction}
    \mu_f^*u_{cl} = \frac{3\sigma}{2\sqrt{2}}\frac{\cos\theta_0-\cos\theta}{\sin\theta} \simeq \mu_f(u_{cl}-u_x) \; ,
\end{equation}
were at the RHS we have the contact line friction coefficient of a model including slip, i.e. where the flow speed at the wall is allowed to deviate from the wall speed. Since we are introducing a non-zero Navier slip length in the CHNS model, we expect the relative slip velocity \textit{at the solid-liquid interface} ($sl$) to be, at a first-order approximation in the shear rate $\dot{\gamma}$:
\begin{equation}    \label{eq:expression-zeta}
    1-\zeta = 1-\frac{\dot{\gamma}\ell_N}{u_w} \simeq 1-\frac{\ell_N}{2H}\; \simeq \frac{u_x}{u_w}\Big|_{sl} \; .
\end{equation}
The values of $\zeta$ (reported in Table~\ref{tab:sharp-interface-slip-md}) are too small to explain the difference between $\mu_f$ and $\mu_f^*$. Conversely, we can determine the relative slip velocity \textit{at the contact line} ($cl$) directly from Equation~\ref{eq:sil-correction} as the ratio of contact line friction coefficients:
\begin{equation}    \label{eq:expression-chi}
    \chi = 1-\frac{\mu_f^*}{\mu_f} \simeq \frac{u_x}{u_w}\Big|_{cl} \; .
\end{equation}
The values of $\chi$ are also reported in Table~\ref{tab:sharp-interface-slip-md}. Figures \ref{fig:md-slip-velocity}b and \ref{fig:md-slip-velocity}c show the collapse of the profiles of the wall-parallel velocity, relative to the wall velocity, obtained by averaging over all the steady simulations. The drops located at the contact lines are quantitatively consistent with the values of $\chi$. Note that the longitudinal velocity drops are consistent with the `apparent perfect slip' occurring at contact lines in steady flows according to to the formulation for the kinematic evolution of the contact angle of \cite{fricke2019angle}.

This analysis suggests that the mismatch between the CHNS-calibrated and MD-measured contact line friction coefficients arises from large slip velocities \textit{localised} at the contact line: this agrees with prior studies reporting significant slip at contact lines \citep{ren2007mdslip}, and with the observed anti-correlation between slip and contact line friction \citep{fernandeztoledano2020md}. Interestingly, this extra slip component \textit{cannot} be attributed to the interface mobility contribution of CHNS, since the dynamic contact angle (and thus the \textit{calibrated} contact line friction coefficient) is insensitive to $M$.

To make our analysis more quantitative, we determine friction coefficients using the CR-GNBC formalism. Following the result by \cite{ren2007mdslip}, we split the liquid/solid friction coefficient $\beta$ into a singular and a regular part, the former localized at the contact line by a Dirac delta function:
\begin{equation}
    \beta = \beta_s\delta_{CL} + \beta_r \; .
\end{equation}
We then smear $\delta_{CL}$ over the contact line region by taking $\delta_{CL}\approx(1-C^2)/(2\sqrt{2}\varepsilon)$; we choose this function as it peaks at the contact line and integrates to 1 when $C$ is the equilibirum hyperbolic tangent solution of CH equations. It is then possible to show that:
\begin{equation}    \label{eq:split-gnbc-friction}
    \beta= \mu_f^*\frac{1-C^2}{2\sqrt{2}\varepsilon} + \Big(1-\frac{u_x}{u_w}\Big)^2\beta_r; .
\end{equation}
The derivation of Equation~\ref{eq:split-gnbc-friction} follows the idea by \cite{fullana2026gnbc} and is reported in the supplementary information. We observe that the localisation term $(1-C^2)$ appears in Equation~\ref{eq:wet-bc}, and this conclude that at the contact line:
\begin{equation}    \label{eq:split-gnbc-friction-muf}
    \mu_f = 2\sqrt{2}\varepsilon\big[1-\chi\big]^2\beta_r + \mu_f^*\; .
\end{equation}
The values of $\beta_r$ estimated from Equation~\ref{eq:split-gnbc-friction-muf} are reported in Table~\ref{tab:sharp-interface-slip-md} in non-dimensional form. The trend traces the one of contact line friction coefficients, with higher values at receding contact lines and on hydrophilic surfaces.

These observation challenges the idea of relying solely on Phase Field diffusion to move contact lines under hydrodynamic no-slip conditions while neglecting contact line friction, as this approach cannot capture contact line asymmetry or local slip variations. While such a protocol can match some results, it still misses some key components engrained in molecular physics.

\section{Conclusions}\label{sec:conclusions}

In this work we have shown a comparison between Molecular Dynamics simulations of a water/hexane Couette flow and a Phase Field model for the motion of the three-phase contact line. We have established a systematic calibration protocol where contact line friction is the sole parameter requiring semi-empirical fitting of MD data. The accurate reconstruction of contact line motion and near-wall interface bending stems by the appropriate choice of contact line friction. The coefficient is obtained by matching the relation between contact line speed and dynamic contact angles observed in MD: the direct transfer of information results in a better agreement between simulation techniques compared to our previous work \citep{lacis2022pellegrino}. This highlights the importance of accounting for both hydrodynamic and molecular sources of energy dissipation in the mesoscale modelling of contact lines. We remark that the parametrization of the CHNS equations is not arbitrary: the determination of CHNS parameters \textit{follows} the calibration of $\mu_f^{\text{adv}}$ and $\mu_f^{\text{rec}}$, and is constrained by numerics and by the sharp interface condition. Subatomic Navier slip is introduced essentially for numerical stability reasons. This demonstrates the robustness of CHNS in describing the physics of contact lines. The possibility of sampling the flow field from MD simulations in both fluid phases enables us to tackle one critical aspect, namely how to reduce the unphysical crossing of the liquid/liquid interface by flow streamlines. The tight adherence to the sharp interface condition reduces crossing, as shown by comparing flow profiles.

Despite its remarkable ability to reproduce several observables, the Phase Field model still misses a slip component, as discussed in Section~\ref{sec:sil-slip-md}. Since constant Navier slip combined with interface diffusion is insufficient, alternative slip models could be explored, e.g. by making the slip length spatially dependent or localized at the interface through its dependence on the order parameter or its gradient. Alternatively, the slip length could be extracted directly from MD simulations following \cite{bocquet2013slip,herrero2019slip}. The Generalized Navier Boundary Condition (GNBC) \citep{qian2003gnbc} provides another option by allowing independent tuning of hydrodynamic slip at the wall and contact line. However, \cite{fullana2026gnbc} have already shown that, at contact lines, the GNBC reduces to a combination of Navier slip and a contact-angle condition similar to Equation~\ref{eq:cl-friciton-sharp-interface-limit}, with the singular friction component being essentially contact line friction. Thus, GNBC is effectively equivalent to the boundary conditions already employed in this study, and an overall improvement is not guaranteed. Indeed, \cite{lacis2020johansson} found that GNBC does not outperform wall relaxation in predicting steady contact-line motion or mitigating streamline crossing.


The omission of the flux coupling term in the present formulation leads to an underestimation of the response of the contact line to large wall velocities, thereby explaining why the Phase Field model progressively deviates from Molecular Dynamics results as the contact line speed increases on hydrophilic surfaces. The inclusion of this term would be expected to improve the agreement at elevated contact line speeds, and constitutes a natural extension of the present model that we leave for future work.

A natural direction for the continuation of this work would involve removing the arbitrariness of the CHNS parameters completely. The main limitation of the method is the choice of $\varepsilon$, which is hard to interpret when considering interfaces with atom-scale sharpness (water/vapour, water/alkanes). Combinations of partially miscible liquids, such as water and alcohol, give rise to a truly diffused interface and thus are good candidates for extending the comparison between MD and CHNS simulations. Furthermore, wider interfaces are characterised by lower surface tension. The CHNS mobility $M$ could then be estimated by simulating interface relaxation, as was for example proposed by \cite{barclay2019mobility}, given that the process would be slow and thus easy to sample. The case of partially miscible biphasic systems however involves some modelling and computational challenges. Partial miscibility is not contemplated in the original formulation of Cahn-Hilliard equations, and the typical double-well potential in the free-energy functional may need to be replaced by a logarithmic Flory-Huggins term. Furthermore, the decrease of surface tension leads to a generally less stable two-phase Couette flow, that needs to be driven with much smaller wall velocities. 

\section*{Supplementary information}
\noindent Supplementary figures and information are available at [...].

\section*{Acknowledgments}
\noindent We thank Prof. Stephane Zaleski (Sorbonne Université) and Dr. Shahab Mirjalili (KTH) for the fruitful scientific discussions and the useful feedback. We also thank Dr. Petter Johannson and Dr. U\v{g}is L\={a}cis for contributing to the flow-binning GROMACS code and to the FreeFem++ code.

\section*{Funding}
\noindent Numerical simulations were performed on resources provided by the Swedish National Infrastructure for Computing (grants NAISS 2023/5-279, NAISS 2023/1-19 and NAISS 2025/1-41) at PDC, Stockholm. We acknowledge funding from the Swedish Research Council (INTERFACE centre grant No. 2016-06119\_VR and grant No. 2021-04820\_VR), from the European Research Council (grant MUCUS StG2019-852529, grant LUBFLOW CoG-101088639 and INTER-ET CoG-2024-101171358).

\section*{Declaration of Interests}
\noindent The authors report no conflict of interest.

\section*{Data availability statement}
\noindent The output of Molecular Dynamics simulations and the configuration files for GROMACS are openly available under Creative Commons Attribution 4.0 International license on Zenodo \citep{zenodohydrophilic,zenodohydrophobic}. The code to perform flow field binning on-the-fly is ported from the GROMACS fork of Dr. Petter Johansson (\href{https://github.com/pjohansson/gromacs-flow-field}{github.com/pjohansson/gromacs-flow-field}). Post-processing scripts are available upon reasonable request. The CHNS FreeFEM++ solver is based on the code by Dr. U\v{g}is L\={a}cis (\href{https://github.com/UgisL/FreeFEM-NS-CH}{github.com/UgisL/FreeFEM-NS-CH}) and is available upon reasonable request. 

\section*{Author ORCID}
\noindent M. Pellegrino, \href{https://orcid.org/0000-0002-2603-8440}{orcid.org/0000-0002-2603-8440}\\
P. K. Kannan, \href{ https://orcid.org/0009-0004-1442-5082}{orcid.org/0009-0004-1442-5082} \\
G. Amberg, \href{https://orcid.org/0000-0003-3336-1462}{orcid.org/0000-0003-3336-1462}\\
S. Bagheri, \href{https://orcid.org/0000-0002-8209-1449}{orcid.org/0000-0002-8209-1449}\\
O. Tammisola, \href{https://orcid.org/0000-0003-4317-1726}{orcid.org/0000-0003-4317-1726}\\
B. Hess, \href{https://orcid.org/0000-0002-7498-7763}{orcid.org/0000-0002-7498-7763}\\

\section*{Author contributions}
\noindent MP, PK, SB, OT and BH conceived the original idea. MP configured the molecular system, carried out Molecular Dynamics simulations, carried out simulations post-processing and compared the results between simulation techniques. PK performed the Phase Field simulations and carried out simulations post-processing. MP and PK wrote the paper with feedback from all coauthors. SB, GA, OT and BH supervised the simulations, the writing and the revision of the manuscript. MP and PK contributed equally to this work.

\appendix

\section{Computation of transport coefficients} \label{sec:app-md}

This appendix summarises the methods used to extract the transport coefficients from molecular dynamics simulations. The density of water and hexane is determined by simple equilibrium simulations at constant temperature and hydrostatic pressure (respectively 300 K and 1 bar). It results $\rho_w=997.9\pm0.1$ kg/m$^3$ and $\rho_h=648.8\pm0.6$ kg/m$^3$, where the uncertainty is due to the fluctuations of the box size when simulating at constant pressure.

The shear viscosity coefficient has been computed both with equilibrium and non-equilibrium simulations. The equilibrium approach consists in evaluating the integral of the off-diagonal components of the pressure tensor of a liquid box at equilibrium, and study its long-time behaviour, according to Einstein's formula:
\begin{equation}    \label{eq:viscosity-einstein}
    \eta = \lim_{t\rightarrow\infty}\frac{1}{6}\frac{V}{k_BT}\frac{d}{dt}\sum_{i\ne j} \expval{\Big(\int_0^tP_{ij}(\tau)d\tau\Big)^2} \; .
\end{equation}
Einstein's formula determines viscosity independently of the shear stress applied to the system and hence it is suitable to characterise simple Newtonian liquids. An alternative approach is to continuously apply a shear deformation to the simulation box and observe the response of the liquid:
\begin{equation}	\label{eq:viscosity-perturbation}
    \eta = P_{xz}\frac{L_z}{U_x} \; ,
\end{equation}
being $U_x$ the maximum deformation velocity at the edge of the simulation box (such that $\dot{\gamma}=U_x/L_x$, being $\dot{\gamma}$ the shear rate). This non-equilibrium approach is more flexible, as it can also be used to study the rheology of non-Newtonian liquids, but it determines the shear viscosity coefficient as a function of the shear rate. Hence, multiple simulations at different shear rates are required: low shear rates produce a noisy but unbiased result, while high shear rates suppress noise but possibly reduce viscosity as the energy input in the system may be too large to be completely dissipated by the thermostat. 

\begin{table}
\begin{center}
\begin{tabular}{p{30mm} | p{23mm} | p{23mm} p{23mm} p{23mm}}
Liquid & Einstein & Deform velocity [nm/ps]   &                    &                    \\
       &          & $U_x=10^{-1}$ & $U_x=10^{-2}$ & $U_x=10^{-3}$ \\ 
       &          &                    &                    &                    \\
SPC/E water    & $0.69\pm0.01$ & $0.69\pm0.01$ & $0.65\pm0.02$ & $0.94\pm0.20$ \\
OPLS-AA hexane & $0.36\pm0.01$ & $0.32\pm0.01$ & $0.36\pm0.04$ & $0.32\pm0.40$
\end{tabular}
\end{center}
\caption{Viscosity in centiPoise [cP] obtained from molecular dynamics simulations using the equilibrium Einstein method and the non-equilibrium shear deformation method, for three different values of the deformation velocity.}
\label{tab:viscosity-md}
\end{table}

Table \ref{tab:viscosity-md} reports the values of shear viscosity obtained from molecular simulations. Equilibrium simulations are approximately 130 ns long and have been repeated 10 times for each liquid. Outputs have been analysed using the approach by \cite{zhang2015viscosity} to extract the viscosity coefficient. Deform runs are 100 ns long and have been repeated for three different deformation velocities. Values obtained with the equilibrium method are chosen as reference for Phase Field simulations due to their low uncertainty and independence from shear rates. The viscosity of hexane agrees well with values obtained both from MD simulations \citep{siu2012oplsaa} and experiments \citep{klein2019viscohex}. The viscosity of water underestimates the experimental value of $\approx0.85$ cP \citep{huber2009water}, as expected for the SPC/E model, but is consistent with previously reported values obtained from MD simulations \citep{gonzales2010viscomdwater}.

\begin{figure}
    \centering
    \includegraphics[width=\linewidth,trim={0 2.85cm 0 2.85cm},clip]{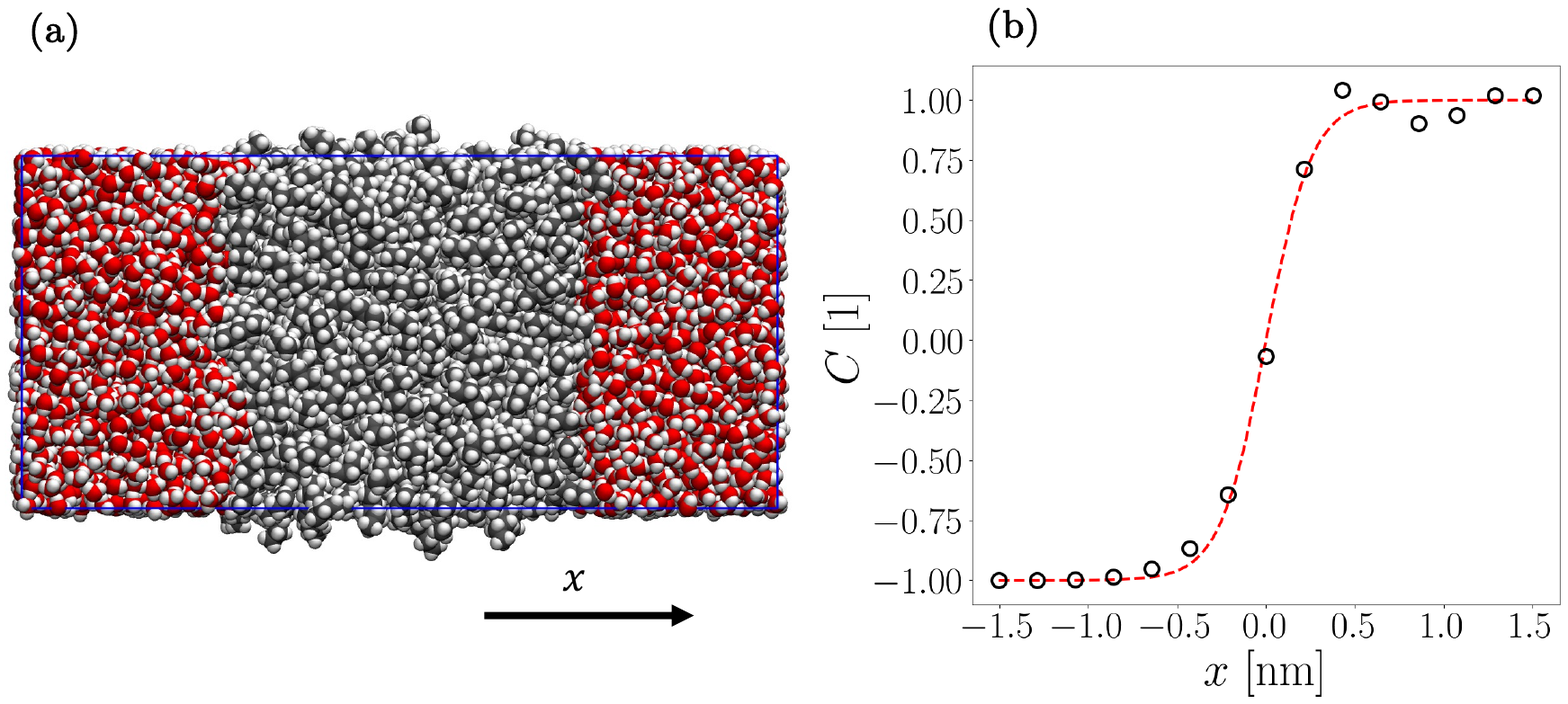}
    \caption{(a): Example of a binary system used to measure water/hexane interface surface tension and width. (b): Interface profile extracted from MD simulations, used to define the intrinsic interface width $\varepsilon_{md}$.}
    \label{fig:md-interface}
\end{figure}

Surface tension is obtained by computing the difference between the component of the pressure tensor that is perpendicular to the water/hexane interface and the average of the components parallel to the interface:
\begin{equation}
    \sigma = \frac{L_z}{2} \Big\{ P_z - \frac{P_x+P_y}{2} \Big\} \; .
\end{equation}
The calculation can be performed on a smaller biphase system without solid walls (fig. \ref{fig:md-interface}). The result is $\sigma=5.19\pm0.24\times10^{-2}$ Pa$\cdot$m, which is in good agreement with previous experiments \citep{goebel1997surftens} and MD simulations \citep{rivera2003surftens}. The same system can be utilised to compute the intrinsic interface width, as defined by for example by \cite{senapati2001interface}. The intrinsic interface width is chosen as reference length instead of the thermal-capillary width as it is independent of the interface area. By fitting the equilibrium solution of the Cahn-Hilliard equation for a one-dimensional interface to the liquid density (averaged along the direction parallel to the interface), we obtain $\varepsilon_{md}=0.18$ nm.

\begin{figure}
    \centering
    \includegraphics[width=\linewidth,trim={0 2cm 0 2cm},clip]{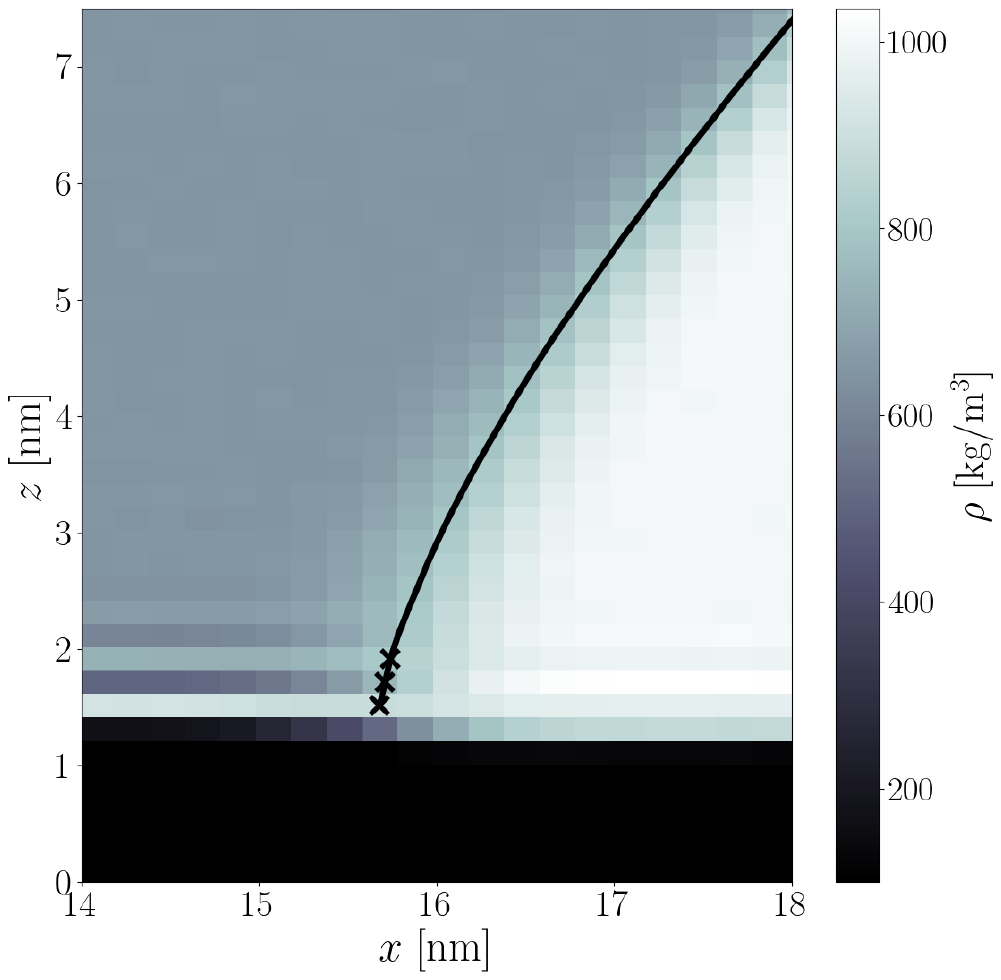}
    \caption{Result of interface extraction for $u_w=3.72$ m/s and $\theta_0=97.3^{\circ}$. The scalar field shows liquid density and the black solid line the liquid/liquid interface. Black crosses close to the contact line indicate the location of the interface points used to compute the dynamic contact angle.}
    \label{fig:md-interface-fit}
 \end{figure}

The horizontal position of the liquid/liquid interface in two-phase Couette flow simulations is determined as a function of the vertical coordinate using a \textit{local} half-density criterion. Essentially, for each row indexed by $j=1,...,N_z$, the average local density of water $\rho_w^{loc}$ is determined by averaging the density field in a section that is far enough from the liquid/liquid interface. The position of the interface is found by scanning across $i=1,...,N_x$ and linearly interpolating between the two bins having density closest to $1/2\cdot\rho_w^{loc}$. This local density criterion allows to have a smooth interface profile even in the near-wall layering region, as it can be observed in Figure~\ref{fig:md-interface-fit}. The density profile of water is chosen to determine the liquid/liquid interface since it's the most dense liquid, and hence the signal-to-noise ratio for the interface position is expected to be larger. Another alternative to the half-density criterion would be to identify the interface where the density of water and the one of hexane are equal. This second criterion however provided noisier measurements of the interface position and of the contact angle. The first bin suitable to extract the liquid/liquid interface is located 0.5 nm away from the reference position of silicon atoms. The equilibrium contact angle $\theta_0$ and the dynamic contact angle $\theta$ are measured respectively by fitting an arc of circle to the interface for $u_{cl}=0$, or by linearly interpolating the first 3 interface points closest to the wall for $u_{cl}\ne0$ and computing the tangent or the cotangent of the slope.


\bibliographystyle{jfm}
\bibliography{references}

\end{document}